\documentclass[12pt,a4paper]{article}

\usepackage{epsfig}
\usepackage{amsmath,amsfonts,amssymb}
\usepackage{cite}
\usepackage{colortbl}
\usepackage{pifont}

\providecommand{\openone}{\leavevmode\hbox{\small1\kern-3.8pt\normalsize1}}

\newcommand{\Tr}{\operatorname{Tr}}
\newcommand{\gp}{g'}

\begin{document}

\begin{center}
\begin{Large}
{\bf Multiboson production in $W'$ decays}
\end{Large}

\vspace{0.5cm}
J.~A.~Aguilar--Saavedra$^{a}$, F.~R. Joaquim$^{b}$ \\[1mm]
\begin{small}
{$^a$ Departamento de F\'{\i}sica Te\'orica y del Cosmos, 
Universidad de Granada, \\ E-18071 Granada, Spain} \\ 
{$^b$ Departamento de F\'{\i}sica and CFTP, Instituto Superior T\'ecnico, Universidade de Lisboa, Lisboa, Portugal} 
\end{small}
\end{center}

\begin{abstract}
In models with an extra $\text{SU}(2)_R$ gauge group and an extended scalar sector, the cascade decays of the $W'$ boson can provide various multiboson signals. In particular, diboson decays $W' \to WZ$ can be suppressed while $W' \to WZX$, with $X$ one of the scalars present in the model, can reach branching ratios around 4\%. We discuss these multiboson signals focusing on possible interpretations of the ATLAS excess in fat jet pair production.
\end{abstract}

\section{Introduction}

A $3.4\sigma$ local excess in boson-tagged jet pair ($JJ$) production reported by the ATLAS Collaboration~\cite{Aad:2015owa}, near an invariant mass $m_{JJ} = 2$ TeV, stands out as the most prominent anomaly that the first run of the Large Hadron Collider (LHC) has left. This excess appears in a dedicated search for heavy resonances decaying into two gauge bosons $WZ$ that subsequently decay hadronically, each boson resulting in one fat jet ($J$). The CMS analysis of the same $JJ$ final state~\cite{Khachatryan:2014hpa} also shows some excess at roughly the same invariant mass. But, intriguingly, complementary searches in the $\ell \nu J$ channel, corresponding to the leptonic decay $W \to \ell \nu$ ($\ell = e,\mu$) and $Z$ hadronic decay, give null results~\cite{Khachatryan:2014gha,Aad:2015ufa}, even if --- as in the case of the ATLAS search --- they are more sensitive to the presence of a resonance. Consequently, the limits from the non-observation of a signal in this decay mode are in tension with the cross section required to explain the excess in ref.~\cite{Aad:2015owa}. The $\ell^+ \ell^- J$ channel with $Z \to \ell^+ \ell^-$ and $W$ decaying hadronically is less sensitive. In the case of the CMS Collaboration~\cite{Khachatryan:2014gha} there is some $\sim 2\sigma$ excess at a smaller invariant mass $m_{\ell \ell J} \sim 1.8$ TeV but the ATLAS analysis~\cite{Aad:2014xka} gives a SM-like result. In addition, heavy resonances decaying into two gauge bosons $VV$ ($V=W,Z$) are also expected to decay into $Vh^0$, with $h^0$ the Higgs boson. Searches for $Vh^0$  in the $JJ$ channel by the CMS Collaboration~\cite{Khachatryan:2015bma} do not show any excess, while a preliminary $Wh^0$ resonance search in the $\ell \nu J$ final state~\cite{CMS:2015gla}, less sensitive than the former, yields a $2.2 \, \sigma$ excess at $m_{Wh^0} = 1.8$ TeV (see ref.~\cite{Aguilar-Saavedra:2015rna} for a detailed discussion).

\begin{figure}[htb]
\begin{center}
\begin{tabular}{ccc}
\includegraphics[height=2.2cm,clip=]{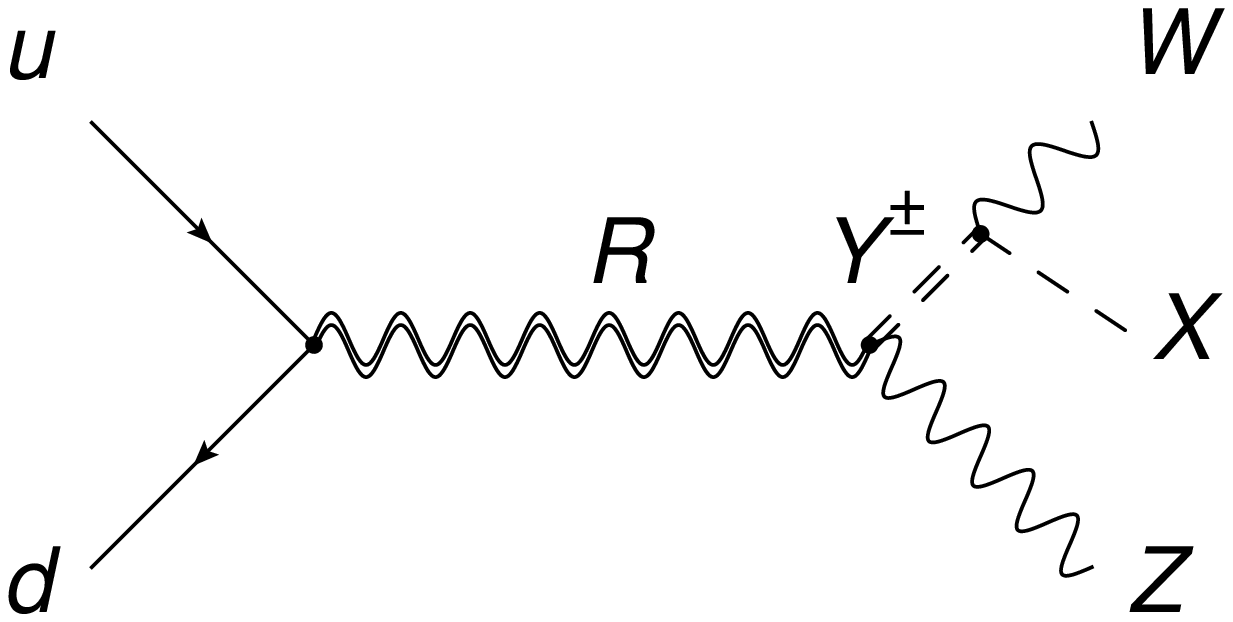} & \quad &
\includegraphics[height=2.2cm,clip=]{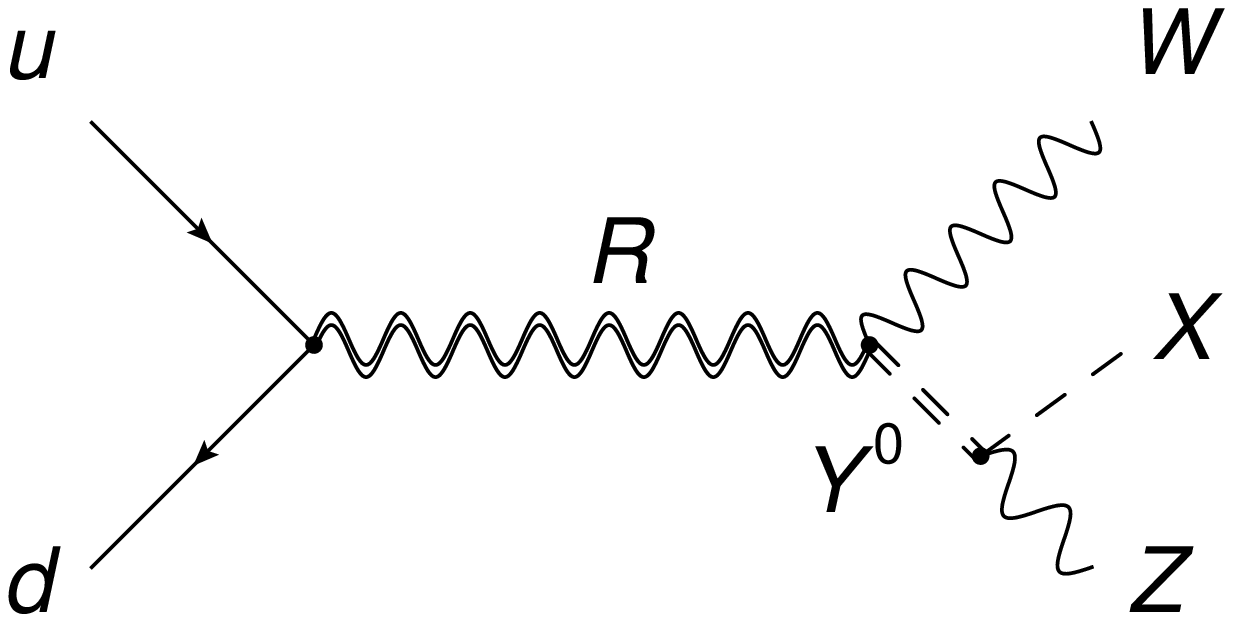}  \\
(a) & & (b) 
\end{tabular}
\caption{Sample diagrams for $R \to VY \to VVX$ production, with $X$ a neutral scalar.}
\label{fig:topVY}
\end{center}
\end{figure}

In order to address the tension between the ATLAS diboson excess~\cite{Aad:2015owa} in the $JJ$ channel and the limits on a possible signal from the other channels~\cite{Khachatryan:2014hpa,Khachatryan:2014gha,Aad:2015ufa,Aad:2014xka}, the hypothesis that this excess is due to diboson production plus an extra particle $X$ was put forward by one of us~\cite{Aguilar-Saavedra:2015rna}. Two production and decay topologies were identified, with a heavy resonance $R$ decaying into $VVX$ via an intermediate on-shell resonance $Y$, as depicted in figure~\ref{fig:topVY}. In both cases, the $VVX$ final state could give a diboson-like signal in the ATLAS analysis~\cite{Aad:2015owa}, while not showing up so conspicuously in the rest of diboson resonance searches. In this paper we present an explicit example of a model where such processes can occur, with $R$ a charged spin-$1$ boson ($W'$), $Y$ a charged ($H^\pm$) or neutral ($H_1^0$) scalar and $X$ a pseudo-scalar ($A^0$) or the Higgs boson ($h^0$). Key ingredients in the model are an additional $\text{SU}(2)'$ gauge group, whose charged member is the $W'$ boson, and an additional scalar doublet to provide the scalars $H^\pm$, $H^0$ and $A^0$.

We note that  many interpretations of the ATLAS excess in terms of a spin-$1$ resonance decaying into $WZ$, $WW$ or $ZZ$ have appeared in the literature~\cite{spin1}, several with an extended scalar sector that couples to $\text{SU}(2)_L$ as well as to a new $\text{SU}(2)'$ gauge group. This is the case, for example, of left-right (LR) models. However, only direct decays $R \to VV$ have been considered, overlooking the tension between the $JJ$ and $\ell \nu J$ analyses or atributing it to statistical
fluctuations.\footnote{Recently, the tension between the $JJ$ excess and the SM-like results in the rest of ATLAS diboson searches has been numerically quantified~\cite{Aad:2015ipg}, and amounts to 2.9 standard deviations. Preliminary results from the second run at 13 TeV leave no significant excess either~\cite{ATLASr2JJ,ATLASr2lnJ,ATLASr2vJ,ATLASr2llJ,CMS:2015nmz}, with a mild $1\,\sigma$ enhancement over the SM prediction near 2 TeV in the ATLAS $JJ$ search.}
Direct $R \to JJ$ decays have also been considered in interpretations  in terms of a new spin-$0$ resonance~\cite{spin0} and other related work~\cite{other}. As we will show in this paper, if the extra scalars present in models with an extra $\text{SU}(2)'$ symmetry group are lighter than the $W'$ boson, their cascade decays can provide multiboson signals.  An alternative explanation of the absence of signals in the $\ell \nu J$ final state is that the diboson excess is due to some new particle having a mass close to the $W$ and $Z$ masses, with hadronic decays, as proposed for example in ref.~\cite{Allanach:2015blv}. Nevertheless, this hypothesis does not explain why a significant excess has not been seen by the CMS Collaboration in their $JJ$ resonance search.

In the remainder of this paper, we will first present in section~\ref{sec:2} the models to be used as a framework. Multiboson $W'$ decays will be discussed in section~\ref{sec:3}, focusing on the dependence of the different (diboson, triboson) signals on the mixing in the scalar sector of the model. The possible multiboson cross sections will be investigated in section~\ref{sec:4}. After this general analysis, we give in section~\ref{sec:5} a couple of benchmark examples where either the triboson signals dominate, or have similar size as diboson signals. We summarise our results in section~\ref{sec:6}.

\section{Framework}
\label{sec:2}

When considering models that can give a $VVX$ signal corresponding to any of the two topologies in figure~\ref{fig:topVY}, we restrict ourselves to particles with spin $0$, $1/2$ or $1$, as those already found in Nature. Furthermore, we consider that $VV=WZ$, since the local significance of the excess with this fat jet selection is larger ($3.4\sigma$)  than for $ZZ$ ($2.9\sigma$) and $WW$ ($2.6\sigma$) selections. In order to reproduce the diboson kinematics, the extra particle $X$ should have a mass $m_X = 100-200$ GeV, and the secondary resonance $Y$ should have a mass below the TeV.

We will assume that the resonance $R$ decaying into $WZX$ is a charge $\pm 1$ particle and $X$ is a neutral one, because a relatively light charged particle $X^\pm$ would be copiously produced in pairs through its gauge coupling to the photon, leading to a dijet pair signal, so far
 unobserved~\cite{ATLAS:2012ds,Chatrchyan:2013izb,Aaltonen:2013hya,Khachatryan:2014lpa}. If $R$ is a heavy $W'$ boson, it would also explain (see for example refs.~\cite{Deppisch:2014qpa,Heikinheimo:2014tba,Aguilar-Saavedra:2014ola}) a $2.8\sigma$ excess in $e^+ e^- jj$ production found by the CMS Collaboration~\cite{Khachatryan:2014dka}, at an invariant mass $m_{eejj} \simeq 2$ TeV. On the other hand, for a charged scalar resonance it is harder to justify the required production cross section (see however ref.~\cite{Chen:2015xql}).
 These arguments motivate us to extend the SM gauge symmetry with an additional $\text{SU}(2)'$.

For the secondary resonance $Y$, the simplest possibility is to have a new scalar. An additional vector boson, perhaps appearing by enlarging the $\text{SU}(2)'$ group, could  yield the production and decay topologies in figure~\ref{fig:topVY} too. However, a lighter gauge boson with a mass of few hundreds of GeV, otherwise undetected, should be (almost) fermiophobic, in contrast with the $W'$ boson resonance produced in the $s$ channel. It is unclear that such possibility is viable. Then, we are led to enlarge the scalar sector of the SM. The mixing of the SM scalar sector with additional $\text{SU}(2)_L$ singlets or triplets is very constrained by Higgs couplings measurements~\cite{Aad:2015pla} and precision electroweak data~\cite{deBlas:2014mba}, therefore we extend the scalar sector with an additional doublet.

The four complex scalar fields in the two $\text{SU}(2)_L$ doublets must transform non-trivially under $\text{SU}(2)'$, in order to couple to the $W'$ boson. It seems more natural to arrange them into two doublets. One possibility is to have a bidoublet, as in LR models; another possibility is that the two $\text{SU}(2)_L$ doublets are $\text{SU}(2)'$ doublets too. We will restrict ourselves to the first option. Also, some of the quark fields must transform non-trivially under $\text{SU}(2)'$, so as to have a $W'$ coupling to quarks. The requirement of gauge invariance of Yukawa terms implies that the $\text{SU}(2)'$ doublets must include right-handed quark fields. In the lepton sector, new neutral leptons $N_R$ can be introduced, embedding the right-handed lepton fields into $\text{SU}(2)'$ doublets. (Alternatively, the $W'$ boson can be leptophobic if the right-handed as well as the left-handed lepton fields are $\text{SU}(2)'$ singlets.) With these assignments, we can identify $\text{SU}(2)'$ with a $\text{SU}(2)_R$ gauge group.

In this work we will discuss two models, which differ in the way the extended gauge group SU(2)$_L\times$SU(2)$_R\times$U(1)$_{B-L}$ is broken to the standard model (SM) one SU(2)$_L\times$U(1)$_{Y}$. We consider two distinct scenarios: the triplet~\cite{Pati:1974yy} and the doublet~\cite{Senjanovic:1978ev} left-right models (TLRM and DLRM, respectively). In the TLRM, the SU(2)$_R\times$U(1)$_{B-L}$ breaking occurs through the vacuum expectation value (VEV) of a $\text{SU}(2)_{R}$ triplet
\begin{equation}
	\Delta_R=\left(\!
	\begin{array}{cc}
		\Delta_R^+/\sqrt{2} &\Delta_R^{++}\\
		\Delta_R^0 &-\Delta_R^+/\sqrt{2}
	\end{array}
	\right)\sim (1,3,2)\;,\;
	\langle \Delta_R \rangle= \frac{1}{\sqrt 2} \left(\!
	\begin{array}{cc}
		0 &0\\
		v_R &0
	\end{array} \!
	\right)\,,
	\label{deltaRdef}
\end{equation}
while in the DLRM, a $\text{SU}(2)_R$ doublet $\chi_R$ is added instead,
\begin{equation}
	\chi_R=\left(\!
	\begin{array}{c}
		\chi^+_R\\
		\chi^0_R
	\end{array}
	\!\right)\sim (1,2,1)\;,\;\langle\chi_R \rangle=\frac{1}{\sqrt 2} \left(\!
	\begin{array}{c}
		0\\
		v_R
	\end{array}
	\!\right)\,.
	\label{def3}
\end{equation}
Gauge interactions of $\Delta_R$ and $\chi_R$ are given by the covariant derivatives
\begin{align}
	D^\mu \Delta_R &= \partial^\mu \Delta_R- i g_R \,\left[\frac{\vec{\tau}}{2} \cdot \vec {W}_{R}^\mu,\Delta_R\right]- i \gp B^\mu \Delta_R\,, \notag \\
	D^\mu \chi_R &= \partial^\mu \chi_R - i g_R \,\frac{\vec{\tau}}{2} \cdot \vec{W}_{R}^\mu \, \chi_R - i\, \frac{\gp}{2}B^\mu \chi_R \,,
	\label{cdtriplet}
\end{align}
where $g_{L,R}$ and $\gp$ are gauge coupling constants. The SU(2)$_{L,R}$ and U(1)$_{B-L}$ gauge fields are denoted by $\vec W_{L,R}^{\mu}$, and $B^\mu$, respectively, and $\vec \tau$ are the Pauli matrices. Notice that we will not impose any discrete symmetry forcing $g_R=g_L$, as in fully LR symmetric models. The SM gauge group is broken down to $\text{U}(1)_{\text{em}}$ by the VEV of a Higgs bidoublet
\begin{equation}
\Phi=\left(
\begin{array}{cc}
\phi_1^0 &\phi_2^+\\
\phi_1^- &\phi_2^0
\end{array}
\right)\sim (2,2,0)\;,\; \tilde{\Phi}=\tau_2 \Phi^\ast \tau_2\,,
\label{Phidef}
\end{equation}
to which corresponds the covariant derivative
\begin{equation}
D^\mu \Phi = \partial^\mu \Phi- i g_{L} \,\vec {W}_{L}^\mu \cdot \frac{\vec{\tau}}{2}\,\Phi+ i g_{R}\, \Phi\,\frac{\vec{\tau}}{2} \cdot \vec {W}_{R}^\mu\,.
\label{def5}
\end{equation}
The vacuum configuration of $\Phi$ is 
\begin{equation}
\langle \Phi \rangle=\frac{1}{\sqrt{2}}\left(
\begin{array}{cc}
v_1 &0\\
0 & e^{i\delta}v_2
\end{array}
\right)\,,
\label{def7}\,
\end{equation}
where $v_1=v\cos\beta$ and $v_2=v\sin\beta$, with $\tan\beta=v_2/v_1$ and $v=246$ GeV. In principle, the phase $\delta$ could trigger spontaneous CP violation in the scalar sector~\cite{Ecker:1983hz}. Although this is an interesting possibility, for the sake of simplicity of our analysis we set $\delta=0$. 

Gauge boson masses and gauge scalar interactions arise from the gauge-invariant scalar kinetic terms:
\begin{align}
\text{TLRM:}\;\mathcal{L}_{\Phi}&=\Tr[(D_\mu \Phi)^\dag(D^\mu \Phi)]+\Tr[(D_\mu \Delta_R)^\dag(D^\mu \Delta_R)] \,, \notag \\
\text{DLRM:}\;\mathcal{L}_{\Phi}&=\Tr[(D_\mu \Phi)^\dag(D^\mu \Phi)]+(D_\mu \chi_R)^\dag D^\mu \chi_R\,.\label{DLRMkin}
\end{align}
The charged gauge boson mass eigenstates are the SM $W$ boson, and a new $W^\prime$ boson, which we identify as being the 2~TeV resonance $R$ in Fig.~\ref{fig:topVY}. From eqs.~(\ref{DLRMkin}) one has, in the limit $v\ll v_R$,
\begin{equation}
M_{W}
\simeq \frac{1}{2}g_L v \left[1-\frac{\epsilon^2}{2k^2}\sin^2 2\beta \right]\;,\; M_{W^\prime} \simeq \frac{k}{2}g_Rv_R\left[1+\frac{\epsilon^2}{2k^2}\right]
\,,
\label{Wmasses}
\end{equation}
where 
\begin{equation}
 \epsilon\equiv \frac{v}{v_R}\simeq k\,\frac{g_R}{\,g_L}\frac{M_W}{M_{W^\prime}}\ll 1
\label{epsdef}\,,
\end{equation}
and $k=\sqrt{2}\,(1)$ for the TLRM (DLRM). In the above equation, the last inequality stems from $ M_{W^\prime} \gg  M_{W}$ and $g_R \sim \mathcal{O}(g_L)$. The mixing between the $W$ and $W^\prime$ mass eigenstates,
\begin{equation}
\left(\! \begin{array}{c}
W_{L}^{\mu}\\
W_{R}^{\mu}
\end{array} \! \right)
= \left(\! \begin{array}{cc}
\cos \zeta & -\sin \zeta \\
\sin \zeta & \cos \zeta
\end{array}\! \right)
\left(\! \begin{array}{c}
W^{\mu}\\
W^{\prime\mu}
\end{array}\! \right)\,,
\label{Wmix}
\end{equation}
is parameterised by an angle $\zeta$ for which
\begin{equation}
\tan 2\zeta= \frac{g_L\,g_R\, \epsilon^2 }{k^2g_R^2+(g_R^2-g_L^2)\epsilon^2}\sin 2\beta \simeq \frac{g_R}{g_L}\frac{M_W^2}{M_{W^\prime}^2}\sin 2\beta \ll 1
\label{Wmixs}\,.
\end{equation}
Except for $W' \to WZ$ decays, which are enhanced by $M_{W'}^2/M_W^2$ due to the longitudinal helicity components, we will neglect $W-W^\prime$ mixing, which is equivalent to considering $W^\pm\sim W_L^\pm$ and $W^{\prime\pm}\sim W_R^\pm$. The physical neutral gauge bosons $Z$, $Z^\prime$ and the photon $A$ are related to the weak SU(2)$_{L,R}$ and U(1)$_{B-L}$ states $W_{L}^{3\mu}$, $W_{R}^{3\mu}$  and $B^\mu$ by 
\begin{equation}
\left(\! \begin{array}{c}
W_{L}^{3\mu}\\
W_{R}^{3\mu}\\
B^\mu
\end{array} \! \right)
\simeq \left(\! \begin{array}{ccc}
c_W & \dfrac{\epsilon^2}{k^4}\cot\theta_W  \cos\varphi \sin^3\varphi& s_W\\
-s_W \cos\varphi & -\sin\varphi& c_W \cos\varphi\\
-s_W \sin\varphi & \cos\varphi& c_W \sin\varphi
\end{array}\! \right) 
\left(\! \begin{array}{c}
Z^{\mu}\\
Z^{\prime\mu}\\
A^\mu
\end{array}\!\right)\,,
\label{Zmix}
\end{equation}
where $c_W\equiv \cos\theta_W$ and $s_W\equiv \sin\theta_W$, $\theta_W$ being the weak mixing angle, and with a new mixing angle $\varphi$ given by
\begin{equation}
\cos\varphi=\frac{g_L}{g_R}\tan\theta_W\,,\quad 
\sin\varphi=\frac{g_L}{\gp}\tan\theta_W \,.
\label{phiangle}\,
\end{equation}
The tangent of the $Z-Z'$ mixing angle $\xi$ is given by the ratio of the $(1,2)$ and $(1,1)$ elements of the mixing matrix,
\begin{equation}
\tan \xi = \frac{\cos \varphi \sin^3 \varphi}{s_W} \frac{\epsilon^2}{k^4} \ll 1 \,.
\end{equation}
At zeroth order in the small parameter $\epsilon$, the mixing between the neutral gauge bosons is completely determined by the requirements that (i) the photon couples to the electric charge; and  (ii) the $Z$ boson couplings to fermions deviate little from the SM prediction. This also sets a relation among the gauge couplings,
\begin{equation}
\gp=\frac{g_Lg_R\tan\theta_W}{\sqrt{g_R^2-g_L^2\tan^2\theta_W}}\,,
\end{equation}
implying $g_R > g_L \tan \theta_W \simeq 0.55 \, g_L$. At zeroth order in $\epsilon$, the masses of the neutral gauge bosons $Z$ and $Z^\prime$ are given by
\begin{equation}
M_{Z} \simeq \frac{g_L v}{2 c_W} \simeq \frac{M_W}{c_W} \,,\quad   M_{Z^\prime} \simeq \frac{k^2}{2} v_R  \sqrt{g_R^2+{\gp}^2} \simeq \frac{k}{\sin \varphi} M_{W'} 
\label{Zmasses}\,.
\end{equation}

In both the TLRM and DLRM, the neutral scalar spectrum contains three CP-even scalars $h^0$ and $H_{1,2}^0$, and one pseudoscalar $A^0$. In the limit $v\gg v_R$ (or equivalently $\epsilon \ll 1$), and barring unnatural cancellations, the neutral complex scalar fields $\phi_{1,2}^0$ and $\Delta_R^0,\chi_R^0$ can be written in terms of the physical fields as
\begin{align}
\phi_1^0 &\simeq \frac{1}{\sqrt{2}}\left[-h^0\sin\alpha +H_1^0 \cos\alpha + i (A^0 \sin\beta+G_1^0\cos\beta)\right] \,, \notag \\
\phi_2^0 &\simeq \frac{1}{\sqrt{2}}\left[h^0\cos\alpha +H_1^0 \sin\alpha + i (A^0 \cos\beta-G_1^0\sin\beta) \right] \,, \notag \\
\Delta_R^0,\chi_R^0 &\simeq \frac{1}{\sqrt{2}}(H_2^0+G_2^0) \label{chir00}\,,
\end{align}
where $G_{1,2}^0$ are the Goldstone bosons and the angle $\alpha$ is the $h_0-H_1^0$ mixing angle, in the notation of the two Higgs doublet model~\cite{Branco:2011iw}. Notice that, in general, $\alpha$ depends on the parameters of the scalar potential (see section~\ref{sec:5}). Moreover, mixing among $h_0, H_1^0$ and $H_2^0$ could also occur. However, and since present experimental results seem to indicate that the properties of $h^0$ are those of the SM Higgs, we will only focus on scenarios which lead to a Higgs mixing pattern like the one given above, with $\alpha$ constrained to lay in the experimentally allowed ranges in the context of a two Higgs doublet model~\cite{Aad:2015pla}.

As for the charged scalar sector, both models include a pair of charged scalars $H^\pm$, which are related to the components of $\Phi$ and $\Delta_R$ (or $\chi_R$) by the relations
\begin{align}
\phi_1^\pm &= \frac{kH^\pm\sin\beta}{\sqrt{k^2+\epsilon^2\cos^2 2\beta }} + G_1^\pm\cos\beta -  \frac{\epsilon\,G_2^\pm \sin\beta \cos 2\beta}{\sqrt{k^2+\epsilon^2\cos^2 2\beta }} \,, \notag \\
\phi_2^\pm &=\frac{kH^\pm\cos\beta}{\sqrt{k^2+\epsilon^2\cos^2 2\beta }} - G_1^\pm\sin\beta-  \frac{\epsilon\,G_2^\pm \cos\beta \cos2\beta }{\sqrt{k^2+\epsilon^2\cos^2 2\beta }} \,, \notag \\
\Delta_R^\pm,\chi_R^\pm &=\frac{\epsilon\,H^\pm}{\sqrt{k^2+\epsilon^2\cos^2 2\beta }} + \frac{kG_2^\pm}{\sqrt{k^2+\epsilon^2\cos^2 2\beta }}\label{chirpm}\,,
\end{align}
where and $G_{1,2}^\pm$ are the charged Goldstone bosons. In the case of the TLRM, there are two doubly-charged scalars $\Delta_R^{\pm\pm}$ that already are physical. Since we are not interested in the phenomenology related with $\Delta_R^{\pm\pm}$, we consider these states to be heavy enough to not play any significant role in our analysis.

In the approximation of eqs.~(\ref{chir00}), and taking $\sqrt{k^2+\epsilon^2\cos^2 2\beta }\simeq k$ in eqs.~(\ref{chirpm}), the relevant couplings between two vector bosons and one scalar are:
\begin{align}
W^+ W^- h^0 [H_1^0] & : \quad g_L M_W \; \sin(\beta-\alpha) \; [ \cos(\beta-\alpha)] \,, \notag \\
W^{\prime\pm} W^\mp h^0 [H_1^0] & : \quad  -g_R M_W \; \cos(\beta+\alpha) \; [\sin(\beta+\alpha)] \,, \notag \\
W^{\prime\pm}W^\mp A^0  &:\quad  \pm i g_R M_W\cos 2\beta \,, \notag \\
Z Z h^0 [H_1^0] &: \quad \frac{g_LM_W}{2c_W^2} \; \sin(\beta-\alpha) \; \left[\cos(\beta-\alpha)\right]\notag \,,\\
Z Z^\prime h^0[H_1^0]  &:\quad  \;\frac{g_RM_W}{c_W}\sin\varphi \; \sin(\beta-\alpha) \; \left[\cos(\beta-\alpha) \right] \,, \notag \\
W^{\prime\pm} Z H^\mp & : \quad - \frac{g_R M_W}{c_W} \cos 2\beta \,.
\end{align}
Notice that the $W^{\prime\pm} Z H^\mp$ interaction receives a contribution from the $\Delta_R$ ($\chi_R$) kinetic term. These contributions differ by a $\sqrt 2$ factor for the triplet and doublet, but the difference is compensated when going to the physical basis, namely eqs.~(\ref{chirpm}).
The relevant couplings of one gauge boson to two scalars are
\begin{align}
W^\pm H^\mp h^0 [H_1^0] &: \quad \mp \frac{g_L}{2} (p_{h^0 [H_1^0]} - p_{H^\pm})^\mu \; \cos(\beta-\alpha) \left[ - \sin(\beta-\alpha) \right] \,, \notag \\
W^\pm H^\mp A^0 &: \quad -i \frac{g_L}{2} (p_{A^0} - p_{H^\pm})^\mu \,, \notag \\
W^{\prime \pm} H^\mp h^0 [H_1^0] &: \quad \mp \frac{g_R}{2} (p_{h^0 [H_1^0]} - p_{H^\pm})^\mu \; \sin(\beta+\alpha) \left[ - \cos(\beta+\alpha) \right] \,, \notag \\ 
W^{\prime \pm} H^\mp A^0 &: \quad i \frac{g_R}{2} (p_{A^0} - p_{H^\pm})^\mu \sin 2 \beta \,, \notag \\ \displaybreak
Z A^0 h^0 [H_1^0] &: \quad - \frac{g_L}{2 c_W} (p_{h^0 [H_1^0]} - p_{A^0})^\mu \; \cos(\beta-\alpha) \left[ - \sin(\beta-\alpha) \right] \,, \notag \\ 
Z' A^0 h^0 [H_1^0] &: \quad - \frac{g_R}{2}  \sin \varphi \, (p_{h^0 [H_1^0]} - p_{A^0})^\mu \; \cos(\beta-\alpha) \left[ - \sin(\beta-\alpha) \right] \,, \notag \\
Z' H^+ H^- &: \quad \frac{g_R}{2} \sin \varphi \, (p_{H^+} - p_{H^-})^\mu \,,
\end{align}
with $p_X^\mu$ the flowing-in four-momentum of particle $X$. 

In both the TLRM and DLRM, the three lepton and quark families are placed in left- and right-handed doublets
\begin{align}
& Q_{Li} = \left(\! \begin{array}{c}
u_i\\
d_i
\end{array} \! \right)_L\sim (2,0,1/3) \,,\quad 
\ell_{Li}= \left(\! \begin{array}{c}
\nu_i\\
e_i
\end{array} \!\right)_L \sim(2,0,-1) \,, \notag \\[1mm]
& Q_{Ri} = \left(\!\begin{array}{c}
u_i\\
d_i
\end{array} \! \right)_R \sim(0,2,1/3) \,,\quad
\ell_{Ri}=\left(\! \begin{array}{c}
\nu_i\\
e_i
\end{array}\! \right)_R \sim (0,2,-1) \,,
\label{def1}
\end{align}
where $i=1,2,3$ is a family index. Gauge interactions among fermions and gauge fields are given by:
\begin{align}
\mathcal{L}_{\text{g}}=\sum_{\psi=Q,\ell} \bar{\psi}_L \gamma^\mu D_{\mu}^L \psi_L + (L\rightarrow R)\;,\; 
D_{\mu}^{L,R}= i\partial_\mu +g_{L,R} \,\frac{\vec{\tau}}{2} \cdot \vec{W}_{\mu L,\mu R}+\, \frac{\gp}{2}B^\mu\,,
\end{align}
while the most general Yukawa Lagrangian is:
\begin{align}
\mathcal{L}_{\text{Yuk}}=-\,\bar{\ell}_L (\,Y_\ell \Phi + \tilde{Y}_\ell \tilde{\Phi})\ell_R
 -\,\bar{Q}_L (\,Y_q \Phi + \tilde{Y}_q \tilde{\Phi})Q_R+\text{h.c.}\,,
 \label{Yuk}
\end{align}
where $Y_{\ell,q}$ and $\tilde{Y}_{\ell,q}$ are general complex Yukawa matrices. In the case of the TLRM, the additional term $-\bar{\ell_R^c} (i\tau_2 \Delta_R)\ell_R$ can be involved in the neutrino mass generation. In general, LRSM models suffer from large flavour-changing neutral current (FCNC) effects due to non-diagonal couplings of the neutral scalars with leptons and quarks. Constraints coming from the analysis of $K_L-K_S$ mass difference require neutral scalar masses larger than $5-10$ TeV~\cite{Mohapatra:1983ae}. This lower bound increases by approximately one order of magnitude if one considers contributions to the CP-violating parameter $\epsilon_{CP}$ coming from $\Delta S=2$ Higgs exchange~\cite{Pospelov:1996fq}. Since in our framework we require that $H_1^0$, $H^\pm$ and $A^0$ are relatively light, the Yukawa interactions given above will, in general, lead to unacceptably large FCNC effects. We will therefore consider that the above couplings are somehow suppressed (perhaps due to some extra symmetry) and fermion masses arise from Yukawa interactions generated, for instance, by higher-order operators. Such possibility has been recently explored in Ref.~\cite{Dobrescu:2015yba}, where a Yukawa pattern of the Type II two Higgs doublet model has been reproduced by considering dimension-6 operators of the type $\bar{\psi}_L \tilde{\Phi}\Delta_R^\dag\Delta_R\psi_R$ and $\bar{\psi}_L \tilde{\Phi}\tilde{\Delta}_R^\dag\tilde{\Delta}_R\psi_R$, with $\tilde{\Delta}_R=\tau_2 \Delta_R \tau_2$. In the DLRM the same reasoning can be applied replacing $\Delta_R$ by the doublet combination $\chi_R\chi_R^\dag$, which transforms as a triplet under $\text{SU}(2)_R$.

\section{$W'$ multiboson decays}
\label{sec:3}

When kinematically allowed, the $W'$ decay widths into two bosons are
\begin{align}
& \Gamma(W' \to WZ) =  \frac{g_R^2}{192 \pi } \sin^2 2\beta \, \frac{\lambda(M_{W'}^2,M_W^2,M_Z^2)^{3/2}}{M_{W'}^5} \notag \\
& \qquad \qquad \qquad \qquad  \times \left( 1 + 10 \frac{M_W^2}{M_{W'}^2} + 10 \frac{M_Z^2}{M_{W'}^2} + \frac{M_W^4}{M_{W'}^4} +  \frac{M_Z^4}{M_{W'}^4} + 10 \frac{M_W^2 M_Z^2}{M_{W'}^4} \right) \,, \notag \\
& \Gamma(W' \to H^\pm Z) = \frac{g_R^2}{192 \pi c_W^2} \cos^2 2 \beta \, \frac{\lambda(M_{W'}^2, M_Z^2, M_{H^\pm}^2)^{1/2}}{M_{W'}} \notag \\
& \qquad \qquad \qquad \qquad \times \left( 1 + 10  \frac{M_Z^2}{M_{W'}^2} - 2  \frac{M_{H^\pm}^2}{M_{W'}^2} + \frac{M_Z^4}{M_{W'}^4} +  \frac{M_{H^\pm}^4}{M_{W'}^4} - 2 \frac{M_Z^2 M_{H^\pm}^2}{M_{W'}^4}
\right) \,, \notag \\
& \Gamma(W' \to WS) = \frac{g_R^2}{192 \pi}x_S^2 \, \frac{\lambda(M_{W'}^2, M_W^2, M_S^2)^{1/2}}{M_{W'}} \notag \\
& \qquad \qquad \qquad \qquad  \times \left( 1 + 10  \frac{M_W^2}{M_{W'}^2} - 2  \frac{M_S^2}{M_{W'}^2} + \frac{M_W^4}{M_{W'}^4} +  \frac{M_S^4}{M_{W'}^4} - 2 \frac{M_W^2 M_S^2}{M_{W'}^4} 
\right) \,, \notag \\
& \Gamma(W' \to H^\pm S) = \frac{g_R^2}{192 \pi } (1-x_S^2)\,  \frac{\lambda(M_{W'}^2, M_{H^\pm}^2,M_S^2)^{3/2}}{M_{W'}^5} \,,
\label{ec:Wdec}
\end{align}
with $x_S^2 = \cos^2 (\beta+\alpha),\, \sin^2 (\beta+\alpha),\, \cos^2 2\beta$ for $S=h^0,H_1^0,A^0$, respectively, and
\begin{equation}
\lambda(x,y,z)=x^2+y^2+z^2-2xy-2xz-2yz \,.
\end{equation}
The partial widths into two fermions are
\begin{align}
& \Gamma(W' \to f \bar f') = \frac{N_c \, g_R^2}{48\pi} \, \frac{\lambda(M_{W'}^2,m_f^2,m_{f'}^2)}{M_{W'}} \notag \\
& \qquad \qquad \qquad \qquad  \times \left( 1 - \frac{m_f^2}{2 M_{W'}^2} - \frac{m_{f'}^2}{2 M_{W'}^2} 
- \frac{m_f^4}{2 M_{W'}^4} - \frac{m_{f'}^4}{2 M_{W'}^4} + \frac{m_f^2 m_{f'}^2}{M_{W'}^4} \right) \,,
\end{align}
with $N_c$ a colour factor.
In the limit that $M_{W'}$ is much larger than the other masses, the branching ratio into two bosons is around $8\%$. 

The scalars $S$ produced in $W'$ decays can further decay into two gauge bosons, a gauge boson plus a lighter scalar, or two fermions. We list here the partial widths, provided the channels are open.  For the decay of the heavy neutral scalar they are
\begin{align}
& \Gamma(H_1^0 \to W W) = \frac{g_L^2}{64 \pi} \cos^2(\beta-\alpha) \frac{M_{H_1^0}^3}{M_W^2}  \left( 1 - 4 \frac{M_W^2}{M_{H_1^0}^2} \right)^{1/2} \left( 1-4 \frac{M_W^2}{M_{H_1^0}^2} + 12 \frac{M_W^4}{M_{H_1^0}^4} \right) \,,
\notag \\ \displaybreak
& \Gamma(H_1^0 \to Z Z) = \frac{g_L^2}{128 \pi c_W^2} \cos^2(\beta-\alpha) \frac{M_{H_1^0}^3}{M_Z^2}  \left( 1 - 4 \frac{M_Z^2}{M_{H_1^0}^2} \right)^{1/2} \left( 1-4 \frac{M_Z^2}{M_{H_1^0}^2} + 12 \frac{M_Z^4}{M_{H_1^0}^4} \right) \,,
\notag \\
& \Gamma(H_1^0 \to Z A^0) = \frac{g_L^2}{64 \pi c_W^2} \sin^2(\beta-\alpha) \, \frac{\lambda(M_{H_1^0}^2,M_Z^2,M_{A^0}^2)^{3/2}}{M_{H_1^0}^3 M_Z^2} \,, \notag \\
& \Gamma(H_1^0 \to H^+ W^-) =  \Gamma(H_1^0 \to H^- W^+) =   \frac{g_L^2}{64 \pi } \sin^2(\beta-\alpha) \, \frac{\lambda(M_{H_1^0}^2,M_W^2,M_{H^\pm}^2)^{3/2}}{M_{H_1^0}^3 M_W^2} \,, \notag \\
& \Gamma(H_1^0 \to f \bar f) =  \frac{N_c h_{ff}^2}{16 \pi} M_{H_1^0} \left( 1 - 4 \frac{m_f^2}{M_{H_1^0}^2} \right)^{3/2} \,,
\label{Hdec}
\end{align}
$f$ being a fermion with Yukawa coupling $h_{ff}$ to $H_1^0$. The $H_1^0 h^0 Z$ coupling vanishes and therefore the decay $H_1^0 \to h^0 Z$ does not take place. The heavy scalar can also decay into $SS = h^0 h^0,A^0 A^0$, with widths
\begin{equation}
\Gamma(H_1^0 \to SS) = \frac{v^2 \lambda_{H_1^0 SS}^2}{32 \pi M_{H_1^0}} \left( 1-4 \frac{m_{S}^2}{M_{H_1^0}^2} \right)^{1/2} \,,
\end{equation}
with $\lambda_{H_1^0 SS}$ dimensionless trilinear couplings of order unity, which depend on the coefficients in the scalar potential (see section~\ref{sec:5}) and the mixing in the scalar sector. We will not consider these decays, which are less important for heavier $H_1^0$. For the pseudoscalar the widths are
\begin{align}
& \Gamma(A^0 \to Z h^0) = \frac{g_L^2}{64 \pi c_W^2} \cos^2(\beta-\alpha) \, \frac{\lambda(M_{A^0}^2,M_Z^2,M_{h^0}^2)^{3/2}}{M_{A^0}^3 M_Z^2} \,, \notag \\
& \Gamma(A^0 \to Z H_1^0) = \frac{g_L^2}{64 \pi c_W^2} \sin^2(\beta-\alpha) \, \frac{\lambda(M_{A^0}^2,M_Z^2,M_{H_1^0}^2)^{3/2}}{M_{A^0}^3 M_Z^2} \,, \notag \\
& \Gamma(A^0 \to H^+ W^-) = \Gamma(A^0 \to H^- W^+) =  \frac{g_L^2}{64 \pi } \,\frac{\lambda(M_{H^\pm}^2,M_W^2,M_{A^0}^2)^{3/2}}{M_{A^0}^3 M_W^2} \,, \notag \\
& \Gamma(A^0 \to f \bar f) =  \frac{N_c (h'_{ff})^2}{16 \pi} M_{A^0} \left( 1 - 4 \frac{m_f^2}{M_{A^0}^2} \right)^{1/2} \,, 
\label{ec:Adec}
\end{align}
with $h'_{ff}$ the Yukawa coupling to $A^0$ of the fermion $f$. For the charged scalar, 
\begin{align}
& \Gamma(H^\pm \to W h^0) = \frac{g_L^2}{64 \pi } \cos^2(\beta-\alpha) \, \frac{\lambda(M_{H^\pm}^2,M_W^2,M_{h^0}^2)^{3/2}}{M_{H^\pm}^3 M_W^2} \,, \notag \\
& \Gamma(H^\pm \to W H_1^0) = \frac{g_L^2}{64 \pi } \sin^2(\beta-\alpha) \, \frac{\lambda(M_{H^\pm}^2,M_W^2,M_{H_1^0}^2)^{3/2}}{M_{H^\pm}^3 M_W^2} \,, \notag \\
& \Gamma(H^\pm \to W A^0) = \frac{g_L^2}{64 \pi } \, \frac{\lambda(M_{H^\pm}^2,M_W^2,M_{A^0}^2)^{3/2}}{M_{H^\pm}^3 M_W^2} \,, \notag \\
& \Gamma(H^\pm \to f \bar f') = \frac{3 h_{ff'}^2}{16\pi} \frac{\lambda(M_{H^\pm}^2,m_f^2,m_{f'}^2)}{M_{H^\pm}} \left( 1 -\frac{m_f^2}{M_{H^\pm}^2} - \frac{m_{f'}^2}{M_{H^\pm}^2} \right) \,,
\end{align}
with $f$, $f'$ two fermions and $h_{ff'}$ their Yukawa coupling to $H^\pm$. The $H^\pm W^\mp Z$ coupling is absent. We remark that the partial widths into two bosons grow with the third power of the mass of the decaying scalar, therefore these decays dominate over the rest of decays as soon as there  is phase space available. Depending on the scalar mass hierarchy, there is a plethora of possible $W'$ cascade decay chains yielding multiboson signals. We will focus on two simple cases: (i) an alignment scenario where $A^0$ is lighter than $H_1^0$ and $H^\pm$; (ii) a small misalignment, and the masses of the three new scalars close so that they decay into SM gauge or Higgs bosons.
Notice that the constraints on a pseudoscalar~\cite{Chatrchyan:2013qga,Khachatryan:2014lpa,Khachatryan:2015tha,Khachatryan:2015baw} are very loose, and greatly depend on the couplings assumed to the different fermions. For the charged scalar, we take a mass safely above current limits~\cite{Agashe:2014kda}, which anyway depend strongly on the parameters of the model. The same applies to the heavy scalar $H_1^0$, which also has suppressed coupling to the $W$ and $Z$ bosons.

\subsection{SM-like Higgs scenario}

\begin{figure}[htb]
\begin{center}
\includegraphics[height=5.2cm,clip=]{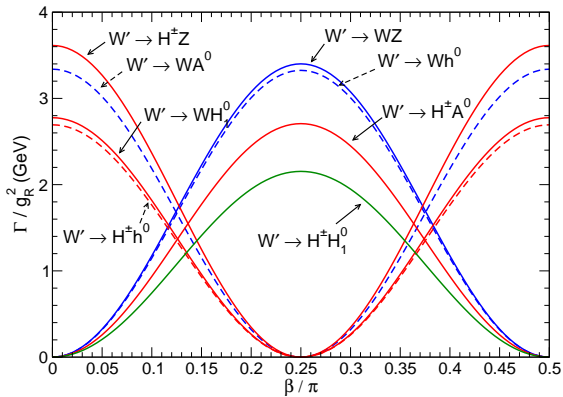}
\caption{$W'$ partial widths into two bosons for the SM-like Higgs scenario. The blue, red and green lines indicate the modes that, upon decays of $H^\pm$ and $H_1^0$, yield dibosons, tribosons and cuadribosons, respectively.}
\label{fig:Wdec-s1}
\end{center}
\end{figure}

We first consider a scenario where $\beta - \alpha= \pi/2$, in which case $h^0$ has the properties of the SM Higgs boson, and with $A^0$ lighter than $H_1^0$ and $H^\pm$, assumed to have equal masses for simplicity. We plot in figure~\ref{fig:Wdec-s1} the partial widths for the  $W'$ decays in eqs.~(\ref{ec:Wdec}), normalised to $g_R=1$, as a function of $\beta$. We take fixed masses $M_{A^0} = 100$ GeV, $M_{H_1^0} = M_{H^\pm} = 500$ GeV. For fixed parameters in the scalar potential, the scalar masses do change with $\beta$, therefore figure~\ref{fig:Wdec-s1} is intended to illustrate the functional dependence on $\beta$ of the different decay widths. (The dependence on the $H^\pm$, $H_1^0$ and $A^0$ masses is due to kinematics, and very mild when they are much lighter than $M_{W'}$.)

\begin{figure}[htb]
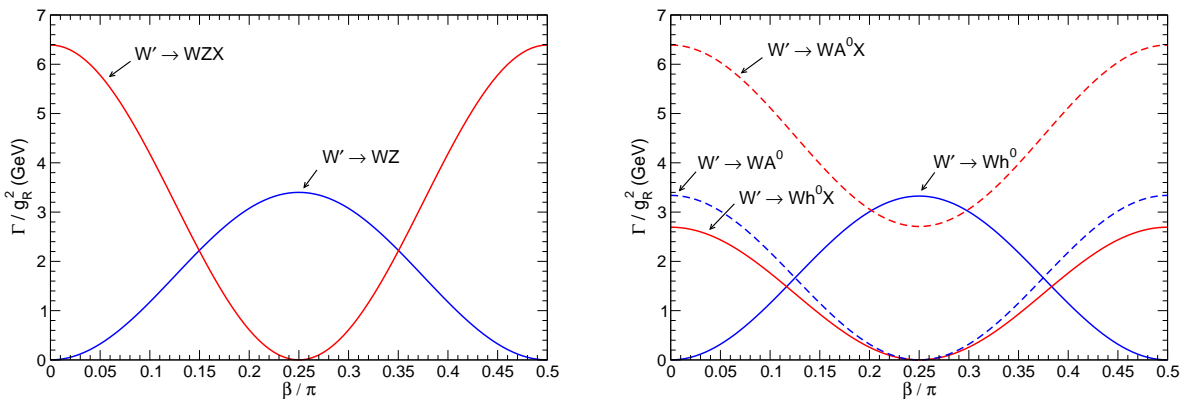

\begin{center}
\begin{tabular}{ccc}
\includegraphics[height=5.2cm,clip=]{Figs/WZvsWZX-s1} & \quad &
\includegraphics[height=5.2cm,clip=]{Figs/WHvsWHX-s1}
\end{tabular}
\caption{Left: $W'$ partial widths into $WZ$ (dibosons) and $WZX$ (tribosons), with $X=A^0$. Right: $W'$ partial widths into $Wh^0$, $WA^0$ (dibosons) and $Wh^0X$, $WA^0X$ (tribosons).}
\label{fig:DvsT-s1}
\end{center}
\end{figure}

In this scenario, the channels $H_1^0 \to Z A^0$ and $H^\pm \to W A^0$ are open and, as aforementioned, these decays are expected to dominate. For example, with the assumed values for the masses, the Yukawa couplings required to have $\Gamma (H_1^0 \to b \bar b / t \bar t) = \Gamma(H_1^0 \to Z A^0)$ are $h_{bb} = 1.04$, $h_{tt}=1.66$, respectively, and the coupling required to have $\Gamma (H^\pm \to t \bar b) = \Gamma(H^\pm \to W A^0)$ is $h_{tb} = 1.2$.
We therefore neglect the decays of $H_1^0$ and $H^\pm$ into quarks, while $A^0$ is expected to decay into $b \bar b$. We collect in table~\ref{tab:sc1} the multiboson signals produced in $W'$ cascade decays, for the scenario here considered.
\begin{table}[htb]
\begin{center}
\begin{tabular}{lclcl}
\multicolumn{1}{c}{dibosons} &\quad & \multicolumn{1}{c}{tribosons} & \quad & \multicolumn{1}{c}{quadribosons} \\
$W' \to WZ$     && $W' \to H^\pm Z \to W A^0 Z$   && $W' \to H^\pm H_1^0 \to W A^0 Z A^0$  \\
$W' \to W h^0$ && $W' \to W H_1^0 \to W Z A^0$ \\
$W' \to W A^0$ && $W' \to H^\pm h^0 \to W A^0 h^0$ \\
                         && $W' \to H^\pm A^0 \to W A^0 A^0$ 
\end{tabular}
\caption{Multiboson signals from $W'$ decays in an alignment scenario with $A^0$ lighter than $H_1^0$ and $H^\pm$.\label{tab:sc1}}
\end{center}
\end{table}
We present in figure~\ref{fig:DvsT-s1} (left) the total size of the $WZ$ diboson (blue) and $WZX$ triboson (red) signals as a function of $\beta$. On the right panel we do the same for the $Wh^0$ and $Wh^0 X$ signals. Additionally, we include the partial widths to $W A^0$ and $W A^0 X$. These final states could mimick the ones with a Higgs boson if $M_{A^0} \sim M_{h^0}$, as the mass window typically used for tagging fat jets as $h^0$ candidates is wide, for example $110 \leq m_J \leq 135$ GeV in ref.~\cite{CMS:2015gla}.

\subsection{Higgs mixing scenario}

\begin{figure}[htb]
\begin{center}
\includegraphics[height=5.2cm,clip=]{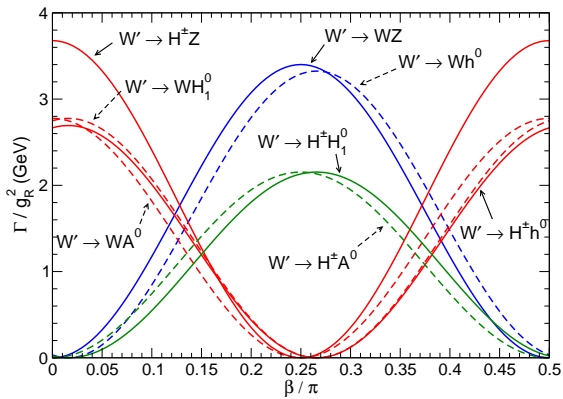}
\caption{$W'$ partial widths into two bosons for the Higgs mixing scenario. The blue, red and green lines indicate the modes that, upon decays of $H^\pm$, $A^0$ and $H_1^0$, yield dibosons, tribosons and cuadribosons, respectively.}
\label{fig:Wdec-s2}
\end{center}
\end{figure}

Current limits on Higgs couplings~\cite{Aad:2015pla} allow for small deviations from the SM prediction,  in particular a small non-zero $\cos (\beta - \alpha)$. We parameterise these deviations introducing a small angle $\gamma$ so that $\beta - \alpha = \pi/2 -\gamma$. We consider a scenario where $H_1^0$, $H^\pm$ and $A^0$ have similar masses so that decays among them are kinematically forbidden (for sufficiently large mass splittings, decays with off-shell $W/Z$ bosons may be important). For simplicity, we take all their masses equal,  $M_{H_1^0} = M_{A^0} = M_{H^\pm} = 500$ GeV. The dependence on the angle $\beta$ of the $W'$ decay widths into two bosons, normalised to $g_R = 1$, is plotted in figure~\ref{fig:Wdec-s2}, taking a small misalignment $\sin \gamma = 0.1$. Notice that there is a small phase shift $\gamma/2$ with respect to figure \ref{fig:Wdec-s1} in the partial widths for $W' \to Wh^0$, $W' \to WH_1^0$, $W' \to H^\pm h^0$, and $W' \to H^\pm H_1^0$.

\begin{figure}[htb]
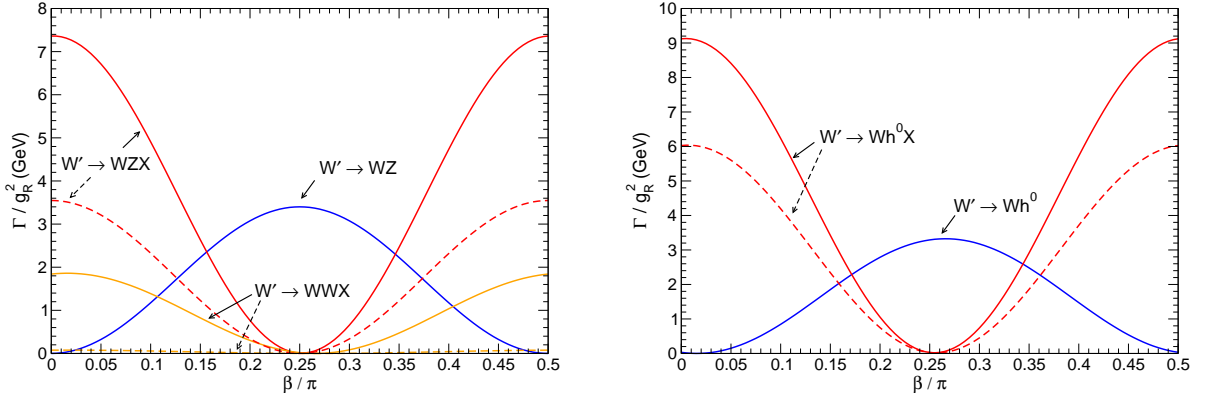

\begin{center}
\begin{tabular}{ccc}
\includegraphics[height=5.2cm,clip=]{Figs/WZvsWZX-s2} & \quad &
\includegraphics[height=5.2cm,clip=]{Figs/WHvsWHX-s2}
\end{tabular}
\caption{Left: $W'$ partial widths into $WZ$ (dibosons) and $WZX$, $WWX$ (tribosons). Right: $W'$ partial widths into $Wh^0$ (dibosons) and $Wh^0X$ (tribosons). For triboson signals, the solid line corresponds to negligible Yukawa couplings and the dashed lines to the assumption given in the text.}
\label{fig:DvsT-s2}
\end{center}
\end{figure}

The small mixing $\cos (\beta - \alpha) = \sin \gamma$ allows decays into SM gauge or Higgs bosons, i.e. $H_1^0 \to W^+ W^-$, $H_1^0 \to ZZ$, $A^0 \to Z h^0$, $H^\pm \to W h^0$, although they compete with the decays into fermions.
We classify in table~\ref{tab:sc2} the possible multiboson signals from $W'$ cascade decays.
\begin{table}[htb]
\begin{center}
\begin{tabular}{lclcl}
\multicolumn{1}{c}{dibosons} &\quad & \multicolumn{1}{c}{tribosons} & \quad & \multicolumn{1}{c}{quadribosons} \\
$W' \to WZ$     &&  $W' \to W H_1^0 \to W W W$   && $W' \to H^\pm H_1^0 \to W h^0 W W$  \\
$W' \to W h^0$ && $W' \to W H_1^0 \to W Z Z$     && $W' \to H^\pm H_1^0 \to W h^0 ZZ$ \\
                         && $W' \to W A^0 \to W Z h^0$  && $W' \to H^\pm A^0 \to W h^0 Z h^0$ \\
                         && $W' \to H^\pm Z \to W h^0 Z$  \\
                         && $W' \to H^\pm h^0 \to W h^0 h^0$ \\
                         
\end{tabular}
\caption{Multiboson signals from $W'$ decays in the Higgs mixing scenario with $H_1^0$, $A^0$ and $H^\pm$ of similar mass, and non-zero $\cos(\beta-\alpha)$.\label{tab:sc2}}
\end{center}
\end{table}
In figure~\ref{fig:DvsT-s2} (left) the total size of the $WZ$ diboson (blue), $WZX$ (red) and $WWX$ (orange) triboson signals is plotted as a function of $\beta$. For triboson signals, the solid lines correspond to negligible Yukawa couplings. For the dashed lines, we have chosen $h_{bb} = h'_{bb}$, equal to the SM bottom quark Yukawa coupling; $h_{tt} = h'_{tt}$, equal to the SM top quark Yukawa coupling; and $h_{tb} = \sqrt{h_{bb} h_{tt}}$.
On the right panel we present the $Wh^0$ and $Wh^0 X$ signals. 

\section{Multiboson cross sections}
\label{sec:4}

So far we have considered the relative size of diboson and triboson signals in two simplified scenarios, and their dependence on the angle $\beta$. We now address the possible size of these signals for a $W'$ boson with a mass near 2 TeV. The next-to-leading order $W'$ cross section~\cite{Duffty:2012rf} at a centre-of-mass (CM) energy of 8 TeV can be parameterised as
\begin{equation}
\sigma_{W'}(\text{pb}) = 638 \, g_R^2 \times \exp \left[ -4.02 M -0.088 M^2 -0.073 M^3 \right] \,,
\end{equation}
with $M$ the $W'$ mass in TeV. The total $W'$ width is nearly independent of $\beta$, $\Gamma = 167 \,g_R^2$ GeV in the alignment scenario and $\Gamma = 166.5 \,g_R^2$ GeV in the Higgs mixing scenario, with a negligible variation of $\pm 0.5$ GeV depending on $\beta$. The approximate $WZ$ diboson and $WZX/WWX$ triboson branching ratios are collected in table~\ref{tab:23BR}.
\begin{table}[htb]
\begin{center}
\begin{tabular}{lccc}
& $\text{Br}(W' \to WZ)$ & $\text{Br}(W' \to WZX)$ & $\text{Br}(W' \to WWX)$ \\
alignment & $0.02 \sin^2 2 \beta$ & $0.039\cos^2 2 \beta$ & $0$ \\
mixing      & $0.02 \sin^2 2 \beta$ & $0.044\cos^2 2 \beta$ & $0.011\cos^2 2 \beta$
\end{tabular}
\caption{Diboson and triboson branching ratios for the Higgs alignment and Higgs mixing scenarios.\label{tab:23BR}}
\end{center}
\end{table}
In both cases we include the decays of the $W'$ boson into the three generations of light leptons plus a heavy neutrino $N$, with a mass taken as 500 GeV.
Notice that in the Higgs mixing scenario the triboson signals may be depleted by the $H_1^0$, $A^0$, $H^\pm$ decays into fermions. The maximum size of diboson plus triboson signals depends on the relative efficiencies of each one, which can only be obtained with a detailed simulation, out of the scope of this work.

The possible size of the coupling $g_R$ is constrained by other processes. Searches for $W' \to t \bar b$ production by the CMS Collaboration yield a limit $\sigma (W' \to t\bar b) \leq 40~\text{fb}$ with a 95\% confidence level (CL)~\cite{Khachatryan:2015edz} for $W'$ masses between 1.9 and 2.2 TeV, where a sum of $t \bar b$ and $\bar t b$ final states is understood. Limits from the ATLAS Collaboration~\cite{Aad:2014xra,Aad:2014xea} are looser.
In a flavour-diagonal scenario (with no $W'$ charged mixing), and independently of the presence of other decay channels, $\Gamma(W' \to WZ) /\Gamma(W' \to t \bar b) \sim \sin^2 2\beta/12$, therefore one has a maximum $\sigma(W' \to WZ) = 3.3$ fb, only one half of the cross section needed to explain the number of excess events at the 2 TeV peak~\cite{Aguilar-Saavedra:2015rna}. Analogously, $\sigma(W' \to WZX+WWX)$ has a maximum of $6-9$ fb, also below the required cross section especially since the efficiency is smaller than for $WZ$. However, the constraint from $W' \to t \bar b$ can be softened or even evaded if a nearly diagonal $W'$ quark mixing matrix is not assumed.

Another constraint results from dijet production. The ATLAS Collaboration sets a limit~\cite{Aad:2014aqa} $\sigma(W' \to jj) \times \mathcal{A} \leq 60$ fb for $M_{W'} = 2$ TeV, with a 95\% CL. With an acceptance $\mathcal{A} \simeq 0.45$~\cite{Aad:2014aqa}, this constraint is translated into $\sigma(W') \leq 280$ fb, i.e. $g_R \leq 1.05$, if all decay channels are open. The CMS Collaboration sets a similar limit~\cite{Khachatryan:2015sja}, $\sigma(W' \to jj) \times \mathcal{A} \leq 100$ fb for $M_{W'} = 2$ TeV. Taking an approximate acceptance of 0.64~\cite{Khachatryan:2015sja} (for isotropic decays) yields a looser limit,  $\sigma(W') \leq 330$ fb. Interestingly, the CMS Collaboration observes a $2 \sigma$ excess but at slightly smaller invariant masses, $m_{jj} \simeq 1.8$ TeV.

\begin{figure}[htb]
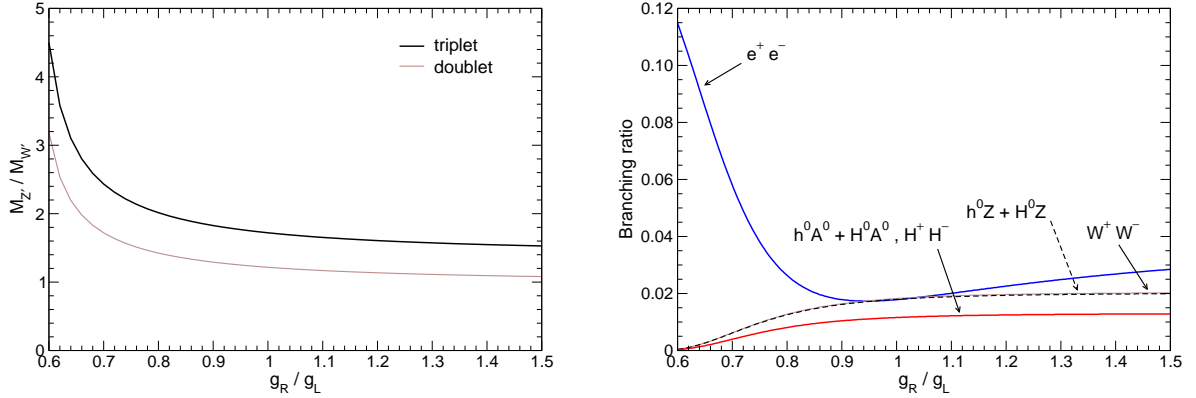

\begin{center}
\begin{tabular}{ccc}
\includegraphics[height=5.2cm,clip=]{Figs/Zpmass} & \quad &
\includegraphics[height=5.2cm,clip=]{Figs/Zdec}
\end{tabular}
\caption{Left: $M_{Z'}/M_{W'}$ ratio as a function of $g_R/g_L$. Right: $Z'$ branching ratio to $e^+ e^-$ and bosonic modes, as a function of $g_R/g_L$}
\label{fig:Zp1}
\end{center}
\end{figure}

A third constraint results from the non-observation of the heavy $Z'$ boson. The relation between the $W'$ and $Z'$ masses depends on the representation of the scalars that break $\text{SU}(2)_R$, and also on the coupling $g_R$. We plot in figure~\ref{fig:Zp1} (left) the ratio $M_{Z'}/M_{W'}$ as a function of $g_R/g_L$ in the two cases that $\text{SU}(2)_R$ is broken by a scalar doublet and a scalar triplet. On the right panel we plot the $Z' \to e^+ e^-$ branching ratio, as well as the branching ratio for the $Z'$ bosonic decay modes, as a function of $g_R/g_L$. The $Z'$ boson is taken much heavier than its decay products. 

\begin{figure}[htb]
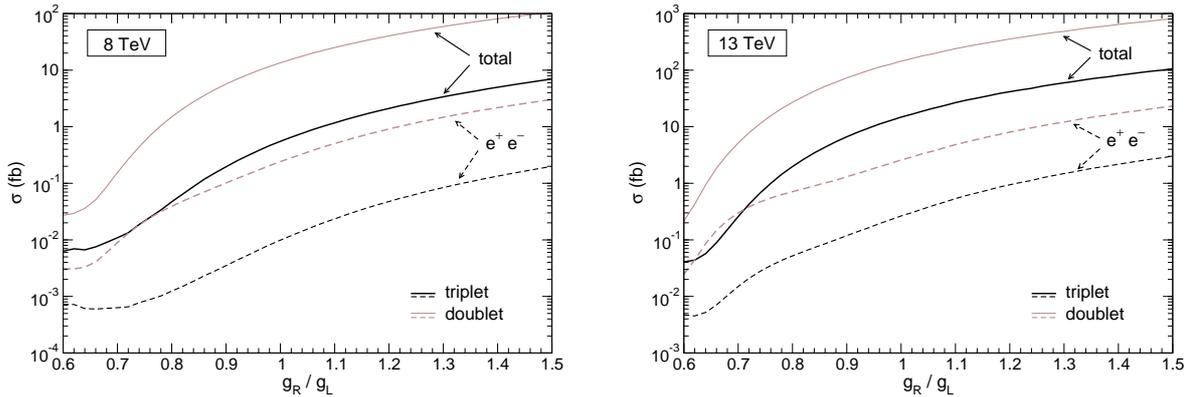

\begin{center}
\begin{tabular}{ccc}
\includegraphics[height=5.2cm,clip=]{Figs/Zprod} & \quad &
\includegraphics[height=5.2cm,clip=]{Figs/Zprod13}
\end{tabular}
\caption{Total $Z'$ production cross section and $Z' \to e^+ e^-$ cross section as a function of $g_R/g_L$, assuming a fixed $W'$ mass of 2 TeV, for CM energies of 8 TeV (left) and 13 TeV (right).}
\label{fig:Zp2}
\end{center}
\end{figure}

Combining the cross section dependence on the mass and couplings, and the coupling dependence of the $Z'$ boson mass, we plot in figure~\ref{fig:Zp2} the total $Z'$ boson production cross section at leading order, as a function of $g_R/g_L$, as well as the $Z' \to e^+ e^-$ cross section, for a reference $W'$ mass of 2 TeV and CM energies of 8 TeV and 13 TeV. A $K$ factor of 1.16~\cite{Aad:2014cka} is included to approximately reproduce the NLO cross section~\cite{Melnikov:2006kv}. For $Z'$ masses of $2-3$ TeV, the unobservation of a signal in the 8 TeV run by the ATLAS Collaboration~\cite{Aad:2014cka} implies $\sigma(Z' \to e^+ e^-) \lesssim 0.2$ fb, assuming lepton universality. Therefore, for a fixed $W'$ mass of 2 TeV, $Z'$ boson searches imply $g_R/g_L \leq 1$ for the doublet, while they do not constrain the range of $g_R$ shown in the case of the triplet. For heavier $W'$ bosons, the limits are looser.

We conclude this section by discussing possible low-energy and precision electroweak data constraints on the $W^\prime$ and $Z^\prime$ masses and mixings~\cite{Langacker:1991zr,Langacker:2008yv,Rizzo:2006nw}. In the specific context of LR symmetric models, limits on $M_{W^\prime}$, $M_{Z^\prime}$ and their corresponding mixing angles have been obtained, for instance, in Refs.~\cite{Langacker:1989xa,Polak:1991pc,Chay:1998hd,Hsieh:2010zr}. For a small $W-W'$ mixing angle $\zeta$ we have, from eq.~(\ref{Wmixs}) and taking $M_{W'} = 2$ TeV,
\begin{equation}
\zeta_g \equiv \frac{g_R}{g_L} \zeta \simeq \frac{1}{2}   \frac{g_R^2}{g_L^2}  \frac{M_W^2}{M_{W'}^2} \sin 2\beta \simeq 0.0008 \, \frac{g_R^2}{g_L^2}  \sin 2\beta \,.
\end{equation}
For $g_R$ of order unity, $\zeta_g$ is below the upper limits in ref.~\cite{Langacker:1989xa}, which are of the order $|\zeta_g| \lesssim (1-2) \times 10^{-3}$, depending on the assumptions about the mixing in the right-handed sector. For the same $W'$ mass, the $Z-Z'$ mixing angle is
\begin{equation}
\xi \simeq \frac{0.0016}{k^2} \frac{(g_R^2-g_L^2 \tan^2 \theta_W)^{3/2}}{g_L g_R^2 c_W} \,,
\end{equation}
with $k^2=2 \; (1)$ for the TLRM (DLRM). The second factor in the above equation is of order unity, {\it e. g.} it is approximately $1.5$ for $g_R=1$, therefore the neutral mixing is compatible with the constraints from low-energy and LEP data, $-0.00040 < \xi < 0.0026$~\cite{Chay:1998hd}. 
A similar analysis presented in ref.~\cite{Hsieh:2010zr} shows that, for $M_{W^\prime} = 2$~TeV, $M_{Z^\prime}/M_{W^\prime} \gtrsim 1.6\,(1.2)$ for the TLRM (DLRM). According to figure~\ref{fig:Zp1}, this implies $g_R/g_L \lesssim 1.2~(1.0)$. Although these bounds are slightly in tension with the cases we are interested in, they can be relaxed with the addition of extra matter content, which naturally appears in embeddings of the $\text{SU}(2)_L \times \text{SU}(2)_R$ in a larger group.\footnote{According to ref.~\cite{Hsieh:2010zr}, the measurements that mainly drive the limits for this model are the $b$-quark forward-backward asymmetry at LEP and the $Z$ boson hadronic width. These two quantities are also modified when vector-like fermions mix with the third generation~\cite{Aguilar-Saavedra:2013qpa}.}

\section{Benchmark examples}
\label{sec:5}

In this section we analyse benchmark scenarios that can account for multiboson production in the context of the TLRM and DLRM, providing some examples of the general behaviour discussed in section~\ref{sec:3}. This requires specifying the scalar potential $V$, which can be written as 
\begin{align}
V=V_\Phi+V_{\Phi_R}+V_{\Phi_R,\Phi}\,,
\label{V}
\end{align}
where $V_\Phi$ and $V_{\Phi_R}$ contain only terms with $\Phi$ and $\Phi_R \equiv \Delta_R$ ($\chi_R$) in the DLRM (TLRM), respectively, and mixed terms involving both $\Phi$ and $\Phi_R$ are included in $V_{\Phi_R,\Phi}$. The most general gauge invariant scalar potential $V_\Phi$ is~\cite{Deshpande:1990ip}
\begin{eqnarray}
	V_\Phi & = & -\mu_1^2 \,\Tr(\Phi^\dag \Phi)-\mu_2^2 \,\left[\Tr(\tilde{\Phi} \Phi^\dag)+\Tr(\tilde{\Phi}^\dag \Phi)\right]+\lambda_1 \Tr(\Phi^\dag \Phi)^2+
	\lambda_2 \left[\Tr(\tilde{\Phi}\Phi^\dag)^2 \right.\nonumber \\
	&& + \left.\Tr(\tilde{\Phi}^\dag\Phi)^2\right] +\lambda_3 \Tr(\tilde{\Phi} \Phi^\dag)\Tr(\tilde{\Phi}^\dag \Phi)+\lambda_4 \Tr(\Phi^\dag \Phi)\left[\Tr(\tilde{\Phi} \Phi^\dag)+\Tr(\tilde{\Phi}^\dag \Phi)\right]\,,
	\label{Vphi}
\end{eqnarray}
where $\mu_i$ are mass parameters, and $\lambda_i$ are dimensionless. (For simplicity we restrict ourselves to the case of real coefficients in the potential.) The pure $\Phi_R$ terms for the TLRM (DLRM) are   
\begin{eqnarray}
V_{\Delta_R}&= &-\mu_3^2\, \Tr(\Delta_R^\dag \Delta_R)+\alpha_1 \Tr(\Delta_R^\dag \Delta_R)^2+\alpha_2 \Tr(\Delta_R \Delta_R) \Tr(\Delta_R^\dag \Delta_R^\dag) \,,\notag\\
V_{\chi_R}&=&-\mu_3^2\, \chi_R^\dag \chi_R+\alpha_1 (\chi_R^\dag \chi_R)^2\label{VCR}\,,
\end{eqnarray}
while for the mixed terms $V_{\Phi_R,\Phi}$ one has
\begin{eqnarray}
V_{\Delta_R,\Phi}&= & \rho_1 \Tr(\Phi^\dag \Phi)\Tr(\Delta_R^\dag \Delta_R)+ \rho_2 \Tr(\Phi^\dag \Phi\Delta_R \Delta_R^\dag) \nonumber\\
 &&+\rho_3\left[\Tr(\Phi\tilde{\Phi}^\dag)+\Tr(\Phi^\dag\tilde{\Phi})\right]\Tr(\Delta_R^\dag \Delta_R)\,, \notag \\
V_{\chi_R,\Phi}&=& \rho_1 \Tr(\Phi^\dag \Phi) \chi_R^\dag \chi_R+ \rho_2\, \chi_R^\dag\Phi^\dag \Phi \chi_R+\rho_3 \chi_R^\dag (\tilde{\Phi}^\dag\Phi+\Phi^\dag\tilde{\Phi})\chi_R\,.
\label{VCRphi}
\end{eqnarray}
Notice that, in general, other invariant dimension-4 combinations of the fields can be included in $V$. However, it can be shown that those can always be written as linear combinations of the terms given above. Detailed analyses of the above potential have been presented in refs.~\cite{
Senjanovic:1978ev,Deshpande:1990ip,Gunion:1989in}. Here, in order to provide representative examples of the benchmark scenarios discussed in section~\ref{sec:3}, it is sufficient to consider simpler cases where some of the parameters in the potential vanish. In the first one, labeled as benchmark A, we impose a discrete symmetry $\Phi\rightarrow i\Phi$ to the scalar potential, and set $v_2=0$~\cite{Senjanovic:1978ev,Gunion:1989in}. This corresponds to having $\mu_2^2=0$ in eq.~(\ref{Vphi}) and $\rho_3=0$ in eqs.~(\ref{VCRphi}). In the second one, labeled as benchmark B, we set $\lambda_4=\rho_1=\rho_3=0$ which, although not motivated by any special symmetry, will allow us to reproduce analytically the Higgs-mixing scenario considered in section~\ref{sec:3}.

In benchmark A the minimisation conditions $\partial V/\partial v_{1,R}=0$ allow to write the mass parameters $\mu_{1,3}^2$ as
\begin{align}
& \mu_1^2=\lambda_1 v^2+\frac{\rho_1}{2}v_R^2 \,,\quad  \mu_3^2=\alpha_1 v_R^2+\frac{\rho_1}{2}v^2\,.
\label{mudefA}
\end{align}
Notice that in this case $v_1=v$ since $v_2=0$. Inserting the above equalities in $V$, one can obtain the neutral scalar masses,
\begin{align}
& m_{h^0}^2\simeq  \frac{4\,\alpha_1\lambda_1-\rho_1^2}{2\alpha_1} \,v^2 \,,
\quad \,\,\,m_{H_1^0}^2\simeq 2\, v^2 (2\lambda_2+\lambda_3)+ \frac{\rho_2}{2}\,v_R^2 \,, \notag \\
& m_{H_2^0}^2\simeq 2 \alpha_1 v_R^2+ \frac{\rho_1^2}{2\alpha_1} v^2 \,,
\quad m_{A^0}^2= \frac{\rho_2}{2}v_R^2 +2v^2(\lambda_3-2\lambda_2) \,,
\label{mH20A}
\end{align}
as well as the charged scalar mass,
\begin{equation}
m_{H^\pm}^2 = \dfrac{\rho_2}{2k^2}\left(k^2v_R^2+v^2\right) \,,
\label{mHpmA}
\end{equation}
where, as before, $k=\sqrt{2}\,(1)$ for the TLRM (DLRM). From these expressions we conclude that $\rho_2$ has to be positive and small in order to yield $m_{H^\pm} \ll v_R$. Also, since we are taking $m_{H_2^0}^2 \sim v_R$, we must have $\alpha_1 >0$. This implies $4\,\alpha_1\lambda_1>\rho_1^2$ to have a positive $m_{h^0}^2$. Inverting these equations, we can obtain approximate expressions for the potential parameters in terms of the scalar masses,
\begin{align}
&\lambda_1\simeq \frac{1}{4v^2}\left( m_{H_1^0}^2+2 m_{h^0}^2-\sqrt{m_{H_1^0}^4-4v_R^2 v^2 \rho_1^2}\right)\;,\; \lambda_2\simeq\frac{m_{H_1^0}^2-m_{A^0}^2}{8v^2}\,,\nonumber\\
&\lambda_3\simeq \frac{1}{4v^2}\left(m_{A^0}^2+m_{H_1^0}^2-2m_{H^\pm}^2\right)\;,\;\alpha_1 \simeq \frac{1}{4v^2}\left( m_{H_1^0}^2+\sqrt{m_{H_1^0}^4-4v_R^2 v^2 \rho_1^2}\right)\,,\nonumber\\ 
&\rho_2\simeq \frac{2m_{H^\pm}^2}{v_R^2}\,.
\end{align}
Choosing a scalar spectrum similar to that considered in the previous section, 
\begin{equation}
m_{h^0}=125~\text{GeV} \,,\quad m_{H_1^0}=m_{H^\pm}=500~\text{GeV} \,, \quad m_{A^0}=100~\text{GeV} \,, \quad m_{H_2^0}=4~\text{TeV} \,,
\label{spect}
\end{equation}
and taking $M_{W^\prime}=2$~TeV and $g_R = 1$, we get for the TLRM and DLRM the parameters
\begin{align}
\text{TLRM}&:\;\lambda_1 \simeq 0.38 \,,\quad \lambda_2 \simeq 0.50 \,,\quad \lambda_3 \simeq -0.98 \,,\quad\alpha_1 \simeq 1.0\,,\quad \rho_2 \simeq 0.06\,, \notag \\
\text{DLRM}&:\;\lambda_1 \simeq 0.63 \,,\quad \lambda_2 \simeq 0.50 \,,\quad \lambda_3 \simeq -0.98 \,,\quad \alpha_1 \simeq 0.50 \,,\quad \rho_2 \simeq 0.03 \label{params}\,,
\end{align}
for $\rho_1=1$. At first order in $\epsilon$ the neutral complex scalar fields $\phi_{1,2}^0$ and $\chi_R^0,\,\Delta_R^0$ can be written as
\begin{align}
& \phi_1^0 \simeq \frac{-h^0  + s_{13} H_2^0 + i G_1^0 }{\sqrt{2}} \,,\quad
\phi_2^0 \simeq \frac{H_1^0 + i A^0  }{\sqrt{2}} \,, \notag \\
& \Delta_R^0,\,\chi_R^0 \simeq \frac{s_{13}h_0 +H_2^0+iG_2^0}{\sqrt{2}}\;, \label{chir0b}
\end{align}
with
\begin{equation}
s_{13} \simeq \frac{\epsilon}{2} \frac{\rho_1}{\alpha_1} \simeq \frac{2\epsilon\rho_1v_R^2}{m_{H_1^0}^2+\sqrt{m_{H_1^0}^4-4\,v_R^2v^2\rho_1^2}} \,.
\end{equation}
When compared with eqs.~(\ref{chir00}), this leads to $\cos(\beta-\alpha)=\cos \alpha=0$, i.e. no Higgs mixing. Notice that the mixing with $H_2^0$ (parameterised by $s_{13}$) is always small, even if $\rho_1 \sim 1$. Besides, in this benchmark the trilinear couplings $\lambda_{H_1^0 h^0 h^0}$ and $\lambda_{H_1^0 A^0 A^0}$ identically vanish.
%
%

In benchmark B, for which $\lambda_4=\rho_1=\rho_3=0$, the minimisation conditions with respect to $v_{1,2}$ and $v_R$ lead to 
\begin{eqnarray}
	\mu_1^2 &= & \lambda_1 v^2 -\frac{\rho_2v_R^2\sin^2\beta}{2\cos 2\beta }\,,\nonumber \\
	\mu_2^2 &= & \left(\lambda_2+\frac{\lambda_3}{2}\right)v^2\sin 2\beta  +\frac{\rho_2}{8}v_R^2\tan 2\beta \,,\nonumber \\
	\mu_3^2 & = & \frac{\rho_2}{2}v^2\sin 2\beta +\alpha_1v_R^2
	\,.
	\label{mudefs}
\end{eqnarray}
The masses of the CP-even and CP-odd scalars are in this case
\begin{eqnarray}
m_{h^0}^2 & \simeq & 2\left[\lambda_1+(2\lambda_2+\lambda_3)\sin^2 2\beta \right]v^2 \,,\notag \\
m_{H_1^0}^2 &\simeq & \frac{\rho_2 v_R^2}{2\cos 2\beta }+2v^2(2\lambda_2+\lambda_3)\cos^2 2\beta \,,\notag \\ 
m_{H_2^0}^2 & \simeq & \frac{v_R^2}{2\alpha_1} \left(4\alpha_1^2+\epsilon^2\,\rho_2^2\sin^4\beta\right)\label{mH202B}\,,\notag \\
m_{A^0}^2 &= & \frac{\rho_2 v_R^2}{2\cos 2\beta }+2v^2(\lambda_3-2\lambda_2)\label{mA0B}
\,,
\end{eqnarray}
where the dependence on the angle $\beta$ is apparent. The charged scalar mass is 
\begin{align}
m_{H^\pm}^2= \frac{\rho_2 }{2k^2 \cos 2\beta } [k^2v_R^2+v^2\cos^2 2\beta ] 
\,.
\end{align}
Again, inverting these equations we can find the potential parameters in terms of the scalar masses,
 \begin{eqnarray}
 \lambda_1 &\simeq & \frac{1}{2v^2}\left[m_{h^0}^2+(m_{H^\pm}^2-m_{H_1^0}^2)\tan^2 2\beta \right]
 \,, \notag \\
 \lambda_2 &\simeq & \frac{1}{8v^2\cos^2 2\beta }\left[m_{H_1^0}^2-m_{H^\pm}^2+(m_{H^\pm}^2-m_{A^0}^2)\cos^2 2\beta \right]\,, \notag \\
 \lambda_3 &\simeq & \frac{1}{4v^2\cos^2 2\beta }\left[m_{H_1^0}^2-m_{H^\pm}^2-(m_{H^\pm}^2-m_{A^0}^2)\cos^2 2\beta \right]\,,\notag \\
 \alpha_1 &\simeq & \frac{m_{H^0_2}^2}{2v_R^2} \,,\quad \rho_2\simeq \frac{2m_{H^\pm}^2\cos 2\beta }{v_R^2}\,.\label{alpha1B}
 \end{eqnarray}
In contrast with benchmark A, here the Higgs mixing pattern is non-trivial. In particular, the alignment condition $\cos(\beta-\alpha)=0$ is not automatically fulfilled since
\begin{equation}
\cos(\beta-\alpha)\simeq  \frac{4\epsilon^2}{\rho_2}  (2\lambda_2+\lambda_3)\cos^2 2\beta
\simeq \frac{\Delta m_H^2+\epsilon^2m_{H_1^0}^2\cos^2 2\beta}{m_{H^\pm}^2}\tan 2\beta\label{cosbma}\,,
\end{equation}
which is still very small if $\Delta m_H^2 \equiv m_{H_1^0}^2-m_{H^\pm}^2=0$. However, by slightly lifting the degeneracy assumption between the $H_1^0$ and $H^\pm$ masses, one can in principle obtain a sizable $h^0-H_1^0$ mixing. Besides, mixing in the $1-3$ and $2-3$ CP-even neutral scalar sectors will be also generated\footnote{Here, $\theta_{13}$ and $\theta_{23}$ are defined according to the standard parameterisation of a unitary $3\times 3$ matrix~\cite{Agashe:2014kda}.},
\begin{align}
\theta_{13}
\simeq
\frac{m_{H^\pm}^4\epsilon \sin\beta\sin(4\beta)}{2 m_{H_2^0}^2(m_{H^\pm}^2- m_{H_2^0}^2)}\;,\;\;
\theta_{23}
\simeq \frac{\epsilon\, m_{H^\pm}^2(m_{H_2^0}^2-m_{H^\pm}^2\sin^2\beta)\cos(2\beta)\sin\beta}{ m_{H_2^0}^2(m_{H_2^0}^2- m_{H^\pm}^2)}\,,
\end{align}
but it is always very small because $\epsilon\ll 1$ and $m_{H^\pm}/m_{H_2}\ll 1$.

As numerical example we take $\cos(\beta-\alpha)\simeq 0.1$ with $\beta=0.1\pi$, in which case $W'$ decays yield diboson plus triboson production. The spectrum is the same as in (\ref{spect}), except for the charged-Higgs mass that we now take as $m_{H^\pm}=530$~GeV, in order to obtain a non-zero Higgs mixing. This spectrum results from the scalar parameters
\begin{align}
\text{TLRM}&:\;\lambda_1 \simeq 0.24 \,,\quad \lambda_2 \simeq 0.47 \,,\quad  \lambda_3 \simeq -1.3 \,,\quad \alpha_1 \simeq 1.0 \,,\quad \rho_2 \simeq 0.06 \,, \notag \\
\text{DLRM}&:\;\lambda_1 \simeq 0.24 \,,\quad  \lambda_2 \simeq 0.47 \,,\quad  \lambda_3 \simeq -1.3 \,,\quad \alpha_1 \simeq 0.50 \,,\quad \rho_2 \simeq 0.03 \label{params2}\,.
\end{align}
We note that in this benchmark we cannot obtain mixing $\cos(\beta-\alpha)\neq 0$ for $\beta = 0$, as it  can be observed from eq.~(\ref{cosbma}).

\section{Discussion}
\label{sec:6}

The ATLAS excess~\cite{Aad:2015owa} in $JJ$ production near $m_{JJ} = 2$ TeV is kinematically compatible with the production of a heavy resonance decaying into two bosons $W/Z$ plus an extra particle $X$, with an intermediate resonance as in figure~\ref{fig:topVY}. As a possible realisation of this mechanism, in this paper we have considered a SM extension with an additional $\text{SU}(2)_R$, in which the new gauge boson $W'$ is the natural candidate to explain the $JJ$ excess.  We have shown in two simple scenarios that, provided the additional scalars present in the model are lighter than the $W'$ boson, the decays $W' \to WZX$ can dominate over decays $W' \to WZ$, as their respective partial widths are proportional to $\cos^2 2 \beta$ and $\sin^2 2 \beta$. In case there is a strong hierarchy among the VEVs of the two neutral scalars that break the $\text{SU}(2)_L$ gauge symmetry, $W' \to WZ$ decays will be largely suppressed ($\sin 2 \beta \sim 0$) with the rate for $W' \to WZX$ reaching its apex ($\cos 2 \beta \sim 1$). If such a hierarchy does not exist, we will have a mixture of $WZ$ and $WZX$ production in general, unless the two VEVs are equal, in which case $WZX$ production is suppressed. The latter is the situation considered in previous literature~\cite{spin1} explaining the excess as $W' \to WZ$ production.

Besides the kinematics, one has to consider the size itself of the observed excess.
For a $\text{SU}(2)_R$ coupling $g_R=1$ and $\cos 2\beta = 1$, the triboson cross section is $\sigma_{WZX} \gtrsim 10$ fb. (For comparison, the maximum diboson signal is one half of this value for the same $g_R$.) While in principle this cross section is of the magnitude needed to explain the excess in ref.~\cite{Aad:2015owa}, the efficiency for triboson signals is expected to be smaller~\cite{Aguilar-Saavedra:2015rna}. A careful evaluation of this efficiency --- which depends not only on the precise details of the boson tagging but also on the identity of the particle $X$ and its mass --- is out of the scope of this work.
In the absence of such a detailed simulation, several qualitative arguments suggest that the efficiency for triboson signals may be not too low so as to explain the ATLAS diboson excess.
\begin{enumerate}
\item The decrease in selection efficiency would be around a factor of six~\cite{Aguilar-Saavedra:2015rna} if only the kinematical configurations where the extra particle $X$ is well separated from the $W$ and $Z$ bosons were to contribute to a ``diboson'' signal after the kinematical selection requirements of the ATLAS analysis~\cite{Aad:2015owa}. However, it is expected that configurations where $X$ (or some of its decay products) merge with the bosons will also contribute to this signal. 
\item In this respect, one of the boson tagging variables used by the ATLAS Collaboration is the jet mass $m_J$, which is required to lie in a suitable interval around the $W$ or $Z$ pole mass. Clearly, if $X$ merges with a boosted $W/Z$ boson, then $m_J$ will increase, thus reducing the boson tagging efficiency compared to the direct $W' \to WZ$ decay. Another tagging variable is the number of tracks $N_\text{trk}$ in the jet, required to be $N_\text{trk} \leq 30$~\cite{Aad:2015owa,ATLASr2JJ}. Likewise, if $X$ merges with a boosted $W/Z$ boson, the number of tracks in the jet will be larger and the boson tagging efficiency will be correspondingly lower. As a consequence of these tagging requirements, for the kinematical configurations where $X$ merges with the $W/Z$ bosons one expects a reduced, but not zero, boson tagging efficiency.
\item In the run 2 $JJ$ search~\cite{ATLASr2JJ}, the ATLAS Collaboration has provided results for the $JJ$ invariant mass distribution when requirements on one of these boson tagging variables are dropped. Interestingly, when the $m_J$ or $N_\text{trk}$ cuts are not applied, slight bumps in the $m_{JJ}$ distributions are seen around $m_{JJ}$ = 2 TeV, which are not visible when the full boson tagging is performed. Although the dataset is still limited by statistics and definite conclusions cannot be drawn, this feature certainly deserves a more detailed investigation.

\item Additional processes may mimick $WZ$ or $WZX$ production, for example $WA^0$, $WA^0A^0$ and $WA^0 h^0$ production, if the new pseudoscalar $A^0$ has a mass similar to the $W/Z$ masses, thus also increasing the potential signal.
\end{enumerate}
On the other hand, the possibility that $g_R$ is larger than unity is in principle allowed, leading to larger triboson cross sections. In this respect, the $W'$ gauge couplings to the quarks can be reduced due to mixing with additional vector-like quarks, as suggested in ref.~\cite{Collins:2015wua}, thereby increasing the $W'$ branching ratios into multiboson final states. (The decrease in $W'$ cross section is compensated by a larger $g_R$.) We also note that direct $W' \to q \bar q$ decays, with the two quarks tagged as boson jets, have also been proposed as additional contributions to the ATLAS $JJ$ excess~\cite{Dobrescu:2015jvn}. By considering the efficiency plots in ref.\cite{Aad:2015owa} and assuming for simplicity that the tagging variables $\sqrt y$, $N_\text{trk}$ and $m_J$ are uncorrelated, we estimate that the tagging efficiency for light jets ($\epsilon_j$) is $1/40$ of the efficiency for true boson jets ($\epsilon_V$). Therefore, the $W' \to q\bar q$ signal will be suppressed by a factor $(\epsilon_j/\epsilon_V)^2 = 1/1600$ and, likely, contributes negligibly to a possible signal. (The $W' \to jj$ signal would be comparable to $W' \to WZ$ if $\epsilon_j=\epsilon_V/5$.) The contribution of $W' \to t\bar b$ with the top and bottom quark jets mistagged as boson jets is expected to be subdominant, because $\sigma(W' \to t \bar b) \leq 40$ fb~\cite{Khachatryan:2015edz}, therefore if we assume a mistagging efficiency $\epsilon_b=\epsilon_V/40$ for $b$-quark jets the possible contribution is marginal.

The ATLAS diboson excess remains an interesting hint for new physics at the LHC, and for sure new run 2 data will bring light on it, settling the issue of whether this peak, if a real effect, is a diboson resonance or something more complex. For a $W'$ mass of 2 TeV, the cross section at 13 TeV is approximately 7 times higher than at 8 TeV, making up for the smaller luminosity alrady collected in 2015. The new measurements at 13 TeV~\cite{ATLASr2JJ,ATLASr2lnJ,ATLASr2vJ,ATLASr2llJ,CMS:2015nmz} are yet inconclusive (although they seem to disfavour the possibility of a $R \to JJ$ resonance), and more data and refined analyses are needed to draw a definite conclusion. Whatever the final outcome of the new measurements is, we have shown in this paper that the scalar sector of models with an extra $\text{SU}(2)_R$ provides a rich variety of multiboson signals that are worth exploring in collider experiments.

\section*{Acknowledgelements}
We thank J. Collins, M. P\'erez-Victoria and J. Santiago for useful comments. This work has been supported by {\em Funda\c{c}\~{a}o para a Ci\^{e}ncia e a Tecnologia} (FCT, Portugal) under the project UID/FIS/00777/2013, by MINECO (Spain) project FPA2013-47836-C3-2-P and by Junta de Andaluc\'{\i}a project FQM 101.


\begin{thebibliography}{99}

\bibitem{Aad:2015owa}
  G.~Aad {\it et al.}  [ATLAS Collaboration],
  arXiv:1506.00962 [hep-ex].

\bibitem{Khachatryan:2014hpa}
  V.~Khachatryan {\it et al.} [CMS Collaboration],
  JHEP {\bf 1408} (2014) 173
  [arXiv:1405.1994 [hep-ex]].

\bibitem{Khachatryan:2014gha}
  V.~Khachatryan {\it et al.} [CMS Collaboration],
  JHEP {\bf 1408} (2014) 174
  [arXiv:1405.3447 [hep-ex]].

\bibitem{Aad:2015ufa}
  G.~Aad {\it et al.} [ATLAS Collaboration],
  Eur.\ Phys.\ J.\ C {\bf 75} (2015) 5,  209
   [Erratum Eur.\ Phys.\ J.\ C {\bf 75} (2015) 370]
  [arXiv:1503.04677 [hep-ex]].

\bibitem{Aad:2014xka}
  G.~Aad {\it et al.} [ATLAS Collaboration],
  Eur.\ Phys.\ J.\ C {\bf 75} (2015) 69
  [arXiv:1409.6190 [hep-ex]].

\bibitem{Khachatryan:2015bma}
  V.~Khachatryan {\it et al.}  [CMS Collaboration],
  arXiv:1506.01443 [hep-ex].

\bibitem{CMS:2015gla}
  CMS Collaboration, 
  Report CMS-PAS-EXO-14-010.

\bibitem{Aguilar-Saavedra:2015rna}
  J.~A.~Aguilar-Saavedra,
  JHEP {\bf 1510} (2015) 099
  [arXiv:1506.06739 [hep-ph]].




\bibitem{spin1}
  J.~Hisano, N.~Nagata and Y.~Omura,
  Phys.\ Rev.\ D {\bf 92} (2015) 5,  055001
  [arXiv:1506.03931 [hep-ph]];
  H.~S.~Fukano, M.~Kurachi, S.~Matsuzaki, K.~Terashi and K.~Yamawaki,
  Phys.\ Lett.\ B {\bf 750} (2015) 259
  [arXiv:1506.03751 [hep-ph]];
  D.~B.~Franzosi, M.~T.~Frandsen and F.~Sannino,
  arXiv:1506.04392 [hep-ph];
  L.~Bian, D.~Liu and J.~Shu,
  arXiv:1507.06018 [hep-ph];
  K.~Cheung, W.~Y.~Keung, P.~Y.~Tseng and T.~C.~Yuan,
  arXiv:1506.06064 [hep-ph];
  Y.~Gao, T.~Ghosh, K.~Sinha and J.~H.~Yu,
  Phys.\ Rev.\ D {\bf 92} (2015) 5,  055030
  [arXiv:1506.07511 [hep-ph]];
  A.~Thamm, R.~Torre and A.~Wulzer,
  arXiv:1506.08688 [hep-ph];
  J.~Brehmer, J.~Hewett, J.~Kopp, T.~Rizzo and J.~Tattersall,
  JHEP {\bf 1510} (2015) 182
  [arXiv:1507.00013 [hep-ph]];
  Q.~H.~Cao, B.~Yan and D.~M.~Zhang,
  arXiv:1507.00268 [hep-ph];
  G.~Cacciapaglia and M.~T.~Frandsen,
  Phys.\ Rev.\ D {\bf 92} (2015) 055035
  [arXiv:1507.00900 [hep-ph]];
  T.~Abe, R.~Nagai, S.~Okawa and M.~Tanabashi,
  Phys.\ Rev.\ D {\bf 92} (2015) 5,  055016
  [arXiv:1507.01185 [hep-ph]];
  J.~Heeck and S.~Patra,
  Phys.\ Rev.\ Lett.\  {\bf 115} (2015) 12,  121804
  [arXiv:1507.01584 [hep-ph]];
  T.~Abe, T.~Kitahara and M.~M.~Nojiri,
  arXiv:1507.01681 [hep-ph];
  A.~Carmona, A.~Delgado, M.~QuirÛs and J.~Santiago,
  JHEP {\bf 1509} (2015) 186
  [arXiv:1507.01914 [hep-ph]];
  H.~S.~Fukano, S.~Matsuzaki and K.~Yamawaki,
  arXiv:1507.03428 [hep-ph];
  L.~A.~Anchordoqui, I.~Antoniadis, H.~Goldberg, X.~Huang, D.~Lust and T.~R.~Taylor,
  Phys.\ Lett.\ B {\bf 749} (2015) 484
  [arXiv:1507.05299 [hep-ph]];
  K.~Lane and L.~Prichett,
  arXiv:1507.07102 [hep-ph];
  A.~E.~Faraggi and M.~Guzzi,
  arXiv:1507.07406 [hep-ph];
  M.~Low, A.~Tesi and L.~T.~Wang,
  Phys.\ Rev.\ D {\bf 92} (2015) 8,  085019
  [arXiv:1507.07557 [hep-ph]];
  P.~Arnan, D.~Espriu and F.~Mescia,
  arXiv:1508.00174 [hep-ph];
  P.~S.~Bhupal Dev and R.~N.~Mohapatra,
  Phys.\ Rev.\ Lett.\  {\bf 115} (2015) 18,  181803
  [arXiv:1508.02277 [hep-ph]];
  A.~Dobado, F.~K.~Guo and F.~J.~Llanes-Estrada,
  arXiv:1508.03544 [hep-ph];
  F.~F.~Deppisch, L.~Graf, S.~Kulkarni, S.~Patra, W.~Rodejohann, N.~Sahu and U.~Sarkar,
  arXiv:1508.05940 [hep-ph];
  U.~Aydemir, D.~Minic, C.~Sun and T.~Takeuchi,
  arXiv:1509.01606 [hep-ph];
  R.~L.~Awasthi, P.~S.~B.~Dev and M.~Mitra,
  arXiv:1509.05387 [hep-ph];
  T.~Li, J.~A.~Maxin, V.~E.~Mayes and D.~V.~Nanopoulos,
  arXiv:1509.06821 [hep-ph].
  P.~Ko and T.~Nomura,
  arXiv:1510.07872 [hep-ph].


\bibitem{Aad:2015ipg}
  G.~Aad {\it et al.} [ATLAS Collaboration],
  arXiv:1512.05099 [hep-ex].

\bibitem{ATLASr2JJ}
ATLAS collaboration,
  ATLAS-CONF-2015-073.

\bibitem{ATLASr2lnJ}
ATLAS collaboration,
  ATLAS-CONF-2015-075.

\bibitem{ATLASr2vJ}
ATLAS collaboration,
  ATLAS-CONF-2015-068.

\bibitem{ATLASr2llJ}
ATLAS collaboration,
  ATLAS-CONF-2015-071.

\bibitem{CMS:2015nmz}
  CMS Collaboration [CMS Collaboration],
  CMS-PAS-EXO-15-002.


\bibitem{spin0}
  C.~W.~Chiang, H.~Fukuda, K.~Harigaya, M.~Ibe and T.~T.~Yanagida,
  JHEP {\bf 1511} (2015) 015
  [arXiv:1507.02483 [hep-ph]];
  G.~Cacciapaglia, A.~Deandrea and M.~Hashimoto,
  Phys.\ Rev.\ Lett.\  {\bf 115} (2015) 17,  171802
  [arXiv:1507.03098 [hep-ph]];
  V.~Sanz,
  arXiv:1507.03553 [hep-ph];
  C.~H.~Chen and T.~Nomura,
  Phys.\ Lett.\ B {\bf 749} (2015) 464
  [arXiv:1507.04431 [hep-ph]];
  Y.~Omura, K.~Tobe and K.~Tsumura,
  Phys.\ Rev.\ D {\bf 92} (2015) 5,  055015
  [arXiv:1507.05028 [hep-ph]];
  W.~Chao,
  arXiv:1507.05310 [hep-ph];
  C.~Petersson and R.~Torre,
  arXiv:1508.05632 [hep-ph];
  C.~H.~Chen and T.~Nomura,
  arXiv:1509.02039 [hep-ph];
  D.~Aristizabal Sierra, J.~Herrero-Garcia, D.~Restrepo and A.~Vicente,
  arXiv:1510.03437 [hep-ph].



\bibitem{other}
  B.~C.~Allanach, B.~Gripaios and D.~Sutherland,
  Phys.\ Rev.\ D {\bf 92} (2015) 5,  055003
  [arXiv:1507.01638 [hep-ph]];
  D.~Kim, K.~Kong, H.~M.~Lee and S.~C.~Park,
  arXiv:1507.06312 [hep-ph];
  S.~P.~Liew and S.~Shirai,
  arXiv:1507.08273 [hep-ph];
  H.~Terazawa and M.~Yasue,
  arXiv:1508.00172 [hep-ph];
  D.~GonÁalves, F.~Krauss and M.~Spannowsky,
  Phys.\ Rev.\ D {\bf 92} (2015) 5,  053010
  [arXiv:1508.04162 [hep-ph]];
  S.~Fichet and G.~von Gersdorff,
  arXiv:1508.04814 [hep-ph];
  L.~Bian, D.~Liu, J.~Shu and Y.~Zhang,
  arXiv:1509.02787 [hep-ph];
  A.~Sajjad,
  arXiv:1511.02244 [hep-ph];
  B.~Bhattacherjee, P.~Byakti, C.~K.~Khosa, J.~Lahiri and G.~Mendiratta,
  arXiv:1511.02797 [hep-ph].






\bibitem{Allanach:2015blv}
  B.~C.~Allanach, P.~S.~B.~Dev and K.~Sakurai,
  arXiv:1511.01483 [hep-ph].


\bibitem{ATLAS:2012ds}
  G.~Aad {\it et al.} [ATLAS Collaboration],
  Eur.\ Phys.\ J.\ C {\bf 73} (2013) 1,  2263
  [arXiv:1210.4826 [hep-ex]].
  
\bibitem{Chatrchyan:2013izb}
  S.~Chatrchyan {\it et al.} [CMS Collaboration],
  Phys.\ Rev.\ Lett.\  {\bf 110} (2013) 14,  141802
  [arXiv:1302.0531 [hep-ex]].

\bibitem{Aaltonen:2013hya}
  T.~Aaltonen {\it et al.} [CDF Collaboration],
  Phys.\ Rev.\ Lett.\  {\bf 111} (2013) 3,  031802
  [arXiv:1303.2699 [hep-ex]].

\bibitem{Khachatryan:2014lpa}
  V.~Khachatryan {\it et al.} [CMS Collaboration],
  Phys.\ Lett.\ B {\bf 747} (2015) 98
  [arXiv:1412.7706 [hep-ex]].

\bibitem{Deppisch:2014qpa}
  F.~F.~Deppisch, T.~E.~Gonzalo, S.~Patra, N.~Sahu and U.~Sarkar,
  Phys.\ Rev.\ D {\bf 90} (2014) 5,  053014
  [arXiv:1407.5384 [hep-ph]].

\bibitem{Heikinheimo:2014tba}
  M.~Heikinheimo, M.~Raidal and C.~Spethmann,
  Eur.\ Phys.\ J.\ C {\bf 74} (2014) 10,  3107
  [arXiv:1407.6908 [hep-ph]].

\bibitem{Aguilar-Saavedra:2014ola}
  J.~A.~Aguilar-Saavedra and F.~R.~Joaquim,
  Phys.\ Rev.\ D {\bf 90} (2014) 11,  115010
  [arXiv:1408.2456 [hep-ph]].

\bibitem{Khachatryan:2014dka}
  V.~Khachatryan {\it et al.}  [CMS Collaboration],
  Eur.\ Phys.\ J.\ C {\bf 74} (2014) 11,  3149
  [arXiv:1407.3683 [hep-ex]].

\bibitem{Chen:2015xql}
  C.~H.~Chen and T.~Nomura in ref.~\cite{spin0}

\bibitem{Aad:2015pla}
  G.~Aad {\it et al.} [ATLAS Collaboration],
  arXiv:1509.00672 [hep-ex].

\bibitem{deBlas:2014mba}
  J.~de Blas, M.~Chala, M.~Perez-Victoria and J.~Santiago,
  JHEP {\bf 1504} (2015) 078
  [arXiv:1412.8480 [hep-ph]].


\bibitem{Pati:1974yy}
J.~C.~Pati and A.~Salam,
Phys.\ Rev.\ D {\bf 10}, 275 (1974)
[Erratum-ibid.\ D {\bf 11}, 703 (1975)];
R.~N.~Mohapatra and J.~C.~Pati,
Phys.\ Rev.\ D {\bf 11}, 2558 (1975);
G.~Senjanovic and R.~N.~Mohapatra,
Phys.\ Rev.\ D {\bf 12}, 1502 (1975).

\bibitem{Senjanovic:1978ev}
G.~Senjanovic,
Nucl.\ Phys.\ B {\bf 153} (1979) 334.

\bibitem{Ecker:1983hz}
G.~Ecker, W.~Grimus and H.~Neufeld,
Nucl.\ Phys.\ B {\bf 247} (1984) 70.

\bibitem{Branco:2011iw}
  G.~C.~Branco, P.~M.~Ferreira, L.~Lavoura, M.~N.~Rebelo, M.~Sher and J.~P.~Silva,
  Phys.\ Rept.\  {\bf 516} (2012) 1
  [arXiv:1106.0034 [hep-ph]].

\bibitem{Mohapatra:1983ae}
R.~N.~Mohapatra, G.~Senjanovic and M.~D.~Tran,
Phys.\ Rev.\ D {\bf 28} (1983) 546;
F.~J.~Gilman and M.~H.~Reno,
Phys.\ Lett.\ B {\bf 127} (1983) 426;
Phys.\ Rev.\ D {\bf 29} (1984) 937;
G.~Ecker and W.~Grimus,
Nucl.\ Phys.\ B {\bf 258} (1985) 328.


\bibitem{Pospelov:1996fq}
M.~E.~Pospelov,
Phys.\ Rev.\ D {\bf 56} (1997) 259.

\bibitem{Dobrescu:2015yba}
B.~A.~Dobrescu and Z.~Liu,
JHEP {\bf 1510} (2015) 118.




















\bibitem{Chatrchyan:2013qga}
  S.~Chatrchyan {\it et al.} [CMS Collaboration],
  Phys.\ Lett.\ B {\bf 722} (2013) 207
  [arXiv:1302.2892 [hep-ex]].


\bibitem{Khachatryan:2015tha}
  V.~Khachatryan {\it et al.} [CMS Collaboration],
  arXiv:1510.01181 [hep-ex].

\bibitem{Khachatryan:2015baw}
  V.~Khachatryan {\it et al.} [CMS Collaboration],
  arXiv:1511.03610 [hep-ex].

\bibitem{Agashe:2014kda}
  K.~A.~Olive {\it et al.} [Particle Data Group Collaboration],
  Chin.\ Phys.\ C {\bf 38} (2014) 090001.

\bibitem{Duffty:2012rf}
  D.~Duffty and Z.~Sullivan,
  Phys.\ Rev.\ D {\bf 86} (2012) 075018
  [arXiv:1208.4858 [hep-ph]].

\bibitem{Khachatryan:2015edz}
  V.~Khachatryan {\it et al.} [CMS Collaboration],
  arXiv:1509.06051 [hep-ex].

\bibitem{Aad:2014xra}
  G.~Aad {\it et al.} [ATLAS Collaboration],
  Eur.\ Phys.\ J.\ C {\bf 75} (2015) 4,  165
  [arXiv:1408.0886 [hep-ex]].

\bibitem{Aad:2014xea}
  G.~Aad {\it et al.} [ATLAS Collaboration],
  Phys.\ Lett.\ B {\bf 743} (2015) 235
  [arXiv:1410.4103 [hep-ex]].

\bibitem{Aad:2014aqa}
  G.~Aad {\it et al.} [ATLAS Collaboration],
  Phys.\ Rev.\ D {\bf 91} (2015) 5,  052007
  [arXiv:1407.1376 [hep-ex]].

\bibitem{Khachatryan:2015sja}
  V.~Khachatryan {\it et al.} [CMS Collaboration],
  Phys.\ Rev.\ D {\bf 91} (2015) 5,  052009
  [arXiv:1501.04198 [hep-ex]].

\bibitem{Aad:2014cka}
  G.~Aad {\it et al.} [ATLAS Collaboration],
  Phys.\ Rev.\ D {\bf 90} (2014) 5,  052005
  [arXiv:1405.4123 [hep-ex]].





\bibitem{Melnikov:2006kv}
  K.~Melnikov and F.~Petriello,
  Phys.\ Rev.\ D {\bf 74} (2006) 114017
  [hep-ph/0609070].

\bibitem{Langacker:1991zr}
  P.~Langacker, M.~x.~Luo and A.~K.~Mann,
  Rev.\ Mod.\ Phys.\  {\bf 64} (1992) 87.

\bibitem{Langacker:2008yv}
  P.~Langacker,
  Rev.\ Mod.\ Phys.\  {\bf 81} (2009) 1199
  [arXiv:0801.1345 [hep-ph]].

\bibitem{Rizzo:2006nw}
  T.~G.~Rizzo,
  hep-ph/0610104.


\bibitem{Langacker:1989xa}
  P.~Langacker and S.~U.~Sankar,
  Phys.\ Rev.\ D {\bf 40} (1989) 1569.

\bibitem{Polak:1991pc}
  J.~Polak and M.~Zralek,
  Nucl.\ Phys.\ B {\bf 363} (1991) 385.

\bibitem{Chay:1998hd}
  J.~Chay, K.~Y.~Lee and S.~h.~Nam,
  Phys.\ Rev.\ D {\bf 61} (2000) 035002
  [hep-ph/9809298].

\bibitem{Hsieh:2010zr}
  K.~Hsieh, K.~Schmitz, J.~H.~Yu and C.-P.~Yuan,
  Phys.\ Rev.\ D {\bf 82} (2010) 035011
  [arXiv:1003.3482 [hep-ph]].

\bibitem{Aguilar-Saavedra:2013qpa}
  J.~A.~Aguilar-Saavedra, R.~Benbrik, S.~Heinemeyer and M.~P\'erez-Victoria,
  Phys.\ Rev.\ D {\bf 88} (2013) 9,  094010
  [arXiv:1306.0572 [hep-ph]].



\bibitem{Deshpande:1990ip}
N.~G.~Deshpande, J.~F.~Gunion, B.~Kayser and F.~I.~Olness,
Phys.\ Rev.\ D {\bf 44} (1991) 837.

\bibitem{Gunion:1989in}
  J.~F.~Gunion, J.~Grifols, A.~Mendez, B.~Kayser and F.~I.~Olness,
  Phys.\ Rev.\ D {\bf 40} (1989) 1546.

\bibitem{Collins:2015wua}
  J.~H.~Collins and W.~H.~Ng,
  arXiv:1510.08083 [hep-ph].
 
\bibitem{Dobrescu:2015jvn}
  B.~A.~Dobrescu and P.~J.~Fox,
  arXiv:1511.02148 [hep-ph].

\end{thebibliography}
\end{document}